**BINARY AND MULTI-CLASS INTRUSION DETECTION IN IOT**

**USING**

**STANDALONE AND HYBRID MACHINE AND DEEP LEARNING MODELS**

by

MD Ahnaf Akif

A Thesis Submitted to the Faculty of

The College of Engineering and Computer Science

In Partial Fulfillment of the Requirements for the Degree of

Master of Science

Florida Atlantic University

Boca Raton, FL

December 2024

**BINARY AND MULTI-CLASS INTRUSION DETECTION IN IOT**

**USING**

**STANDALONE AND HYBRID MACHINE AND DEEP LEARNING MODELS**

by

Md Ahnaf Akif

This thesis was prepared under the direction of the candidate's thesis advisor, Dr. Imadeldin Mahgoub, Department of Electrical Engineering and Computer Science, and has been approved by all members of the supervisory committee. It was submitted to the faculty of the College of Engineering and Computer Science and was accepted in partial fulfillment of the requirements for the degree of Master of Science.

SUPERVISORY COMMITTEE:

Imadeldin Mahgoub, Ph.D.
Thesis Advisor

Mohammad Ilyas, Ph.D.

Waseem Asghar, Ph.D.

Hari Kalva, Ph.D.
Chair, Department of Electrical
Engineering and Computer Science

Stella Batalama, Ph.D.
Dean, College of Engineering and Computer Science

Robert W. Stackman Jr., Ph.D.
Dean, Graduate College

November 15, 2024
Date

# ACKNOWLEDGEMENTS

This work is supported by the Office of the Secretary of Defense (OSD) Grant number W911NF2010300.

Almighty Allah, the greatest of all, is first and foremost than anybody else; it is to him that I turn for guidance and strength. Allah has blessed me with the opportunity, determination, and strength to pursue this research. Throughout my life and even more during the period of my study, his constant grace and kindness accompanied me.

I would want to especially thank my adviser, Dr. Imad Mahgoub for his continual support during my MSc, his tolerance, encouragement, and great expertise. His direction constantly helped me with my dissertation writing and research. Having a better adviser and mentor for my MSc would have been unthinkable to me. I owe Allah enough for giving such a talented and generous professor like Dr. Mahgoub as my advisor who had helped me as a friend in this crucial period of my career. In addition to being an invaluable resource, he played a key role in developing my research skills; I would love to have the opportunity to learn even more from him.

In addition to my adviser, I would like to thank the other members of my thesis committee, Dr. Mohammad Ilyas and Dr. Waseem Asghar, for their helpful feedback and support during the entire process. I'm also grateful to them for asking the questions that prompted me to look at my research from different perspectives. Each and every one of them deserves my eternal gratitude for reading my progress reports throughout the semester and providing feedback that has allowed me to refine my work for each subsequent edition.

Additionally, I'd want to thank Dr. Ismail Butun and Andre Williams, who were both accessible and knowledgeable throughout the final six months of my thesis, for their assistance and support. I greatly appreciated their suggestions.

In closing, I want to thank my parents, Nilufar Akter and MD Mominul Haque, for all the love and support they have given me. I am eternally grateful to them for all the spiritual and financial assistance they have given me and the prayers they have offered me during this time.

# ABSTRACT

Author:             Md Ahnaf Akif.

Title:              Binary and Multi Class Intrusion Detection in IoT
                    Using Standalone and Hybrid Machine and Deep
                    Learning Models

Institution:        Florida Atlantic University

Thesis Advisor:     Dr. Imadeldin Mahgoub

Degree:             Master of Science.

Year:               2024

Maintaining security in IoT systems depends on intrusion detection since these networks' sensitivity to cyber-attacks is growing. Based on the IoT23 dataset, this study explores the use of several Machine Learning (ML) and Deep Learning (DL) along with the hybrid models for binary and multi-class intrusion detection. The standalone machine and deep learning models like Random Forest (RF), Extreme Gradient Boosting (XGBoost), Artificial Neural Network (ANN), K-Nearest Neighbors (KNN), Support Vector Machine (SVM), and Convolutional Neural Network (CNN) were used. Furthermore, two hybrid models were created by combining machine learning techniques: RF, XGBoost, AdaBoost, KNN, and SVM and these hybrid models were voting based hybrid classifier. Where one is for binary, and the other one is for multi-class classification. These models were tested using precision, recall, accuracy, and F1-score criteria and compared the performance of each model. This work thoroughly explains how hybrid, standalone ML and DL techniques could improve IDS (Intrusion Detection System) in terms of accuracy and scalability in IoT (Internet of Things).

**BINARY AND MULTI-CLASS INTRUSION DETECTION IN IOT**

**USING**

**STANDALONE AND HYBRID MACHINE AND DEEP LEARNING MODELS**

# LIST OF TABLES

# LIST OF FIGURES

# CHAPTER 1
# INTRODUCTION

The fast-expanding Internet of Things (IoT) device industry offers new challenges for network security. As more linked devices are employed, these networks have become quite sensitive to numerous types of cyber-attacks; so, powerful Intrusion Detection Systems (IDS) have to be designed to safeguard IoT settings. Although helpful in some cases, conventional IDS systems struggle to handle the complexity and variety of modern cyber-attacks, particularly in IoT networks, which usually have resource-constrained devices. This has motivated research of sophisticated ML and DL approaches for both binary (attack vs. normal) and multi-class (various kinds of assaults) intrusion detection.

Recent research on intrusion detection in network traffic data has demonstrated the efficiency of numerous machine learning techniques like K-Nearest Neighbors (KNN) XGBoost, Random Forest, and Support Vector Machine (SVM) [1]. We have seen promising outcomes in detecting complex patterns in invasion data using deep learning approaches, like as Convolutional Neural Networks (CNN) and, Artificial Neural Networks (ANN) especially when dealing with large datasets like the IoT23 dataset [2]. However, when handling multi-class classification or sophisticated attack situations, the performance of standalone models can occasionally be constrained.

Recent work suggests a hybrid strategy combining the strengths of many machine learning techniques to handle these difficulties. This work specifically combines two hybrid models—one for binary intrusion detection and another for multi-class intrusion detection—random forest, XGBoost, AdaBoost, KNN, and SVM. Evidence from earlier research suggests that hybrid models can increase detection accuracy by combining the strengths of several techniques, since they beat independent classifiers in all three metrics[3]. Training and testing the models came from the IoT23 dataset, especially intended for IoT-based intrusion detection research. From Denial of Service (DoS) to distributed attacks, this dataset offers a large supply of data for spotting different kinds of attacks, so acting as a useful benchmark for binary and multi-class intrusion detection research. With an eye toward how effectively they manage class imbalance problems typically observed in IoT datasets, the hybrid models were assessed using key performance standards including precision, F1-score, accuracy, and recall.

Furthermore, by using the capabilities of several classifiers, hybrid models combining these machine learning models especially via voting systems offer even further improvements in detection accuracy [4].

Finally, this work evaluates stand-alone machine learning, deep learning, and hybrid models in terms of precision, F1-score, accuracy, and recall. The results of this work provide interesting study of the development of stronger and scalable IDS solutions able to meet the evolving scene of cyber-attacks in IoT systems.

## 1.1 Motivation and Problem Statement

The fast spread of IoT networks over many sectors has presented major security concerns because of the sheer quantity of linked devices and the changing nature of cyber threats. Often inadequate for properly controlling the complexity and variety of threats in these settings, traditional intrusion detection systems (IDS) were made but these IDS systems have limitations because the new IoT devices has more complexity so, the demand for sophisticated, dependable, and scalable intrusion detection techniques has never been more pressing.

Big data, ML, and DL techniques reveal intriguing approaches to enhance IDS by pointing up dangerous patterns. RF, XGBoost, AdaBoost, KNN, SVM, and DL models, including ANN and CNN, have helped binary and multi-class intrusion detection tremendously. As these standalone models have some limitations, two hybrid models combining the strengths of these stand-alone models have also been tested in the complicated IoT context.

Improving the efficiency of both independent ML and DL models and hybrid models that integrate the best features of many models into a single, voting-based model is the driving force behind this effort. As the IoT devices and attacks on those devices are increasing proportionally, detecting these attacks is becoming much more challenging. Therefore, in order to address this issue, it is necessary to design an intrusion detection system that is both strong and scalable. This is the primary objective of this work.

## 1.2 Contribution

The contributions of this thesis are organized as follows:

- Pre-processing steps of IoT23 and the process includes feature scaling, feature engineering, categorical feature conversion and missing data handling. (Chapter 3)

- Binary classification of IoT attacks using standalone ML and DL models such as RF, XGBoost, ANN and CNN. (Chapter 4)

- Multi class intrusion detection of IoT attacks using standalone ML and DL models such as RF, XGBoost, ANN and CNN. (Chapter 5)

- Hybrid approach for detecting IoT attacks using a voting classifier where different machine learning models are combined. (Chapter 6)

# CHAPTER 2

# LITERATURE REVIEW

Given their growing and linked character, IoT networks—which are increasingly vulnerable to cyber-attacks—rely on intrusion detection systems (IDS) to maintain their security. Using both binary and multi-class intrusion detection models, several machine ML and DL methods have been employed to improve IDS. This review of the literature investigates several current works addressing intrusion detection problems using various techniques, both as stand-alone models and in hybrid setups.

The authors of [5] presented a new AI framework that detects malware in IoT devices to mitigate cyber-attacks. The authors focus on enhancing security in various use cases for smart environments through an all-inclusive AI-enabled approach in this paper. Emulation of a smart environment employing the Raspberry Pi and NVIDIA Jetson as gateways in capturing data from IoT devices connected via the MQTT protocol, therefore enabling monitoring of real-time malware attacks for their prediction. In this work, many models of AI have been evaluated, among which the DNN model demonstrated superior accuracy and classification capability with an F1-score of 92% and detection accuracy of 93% on Edge-IIoTset and IoT-23. Concerns about the impact on system resources by specifying metrics are drawn to traffic and CPU usage on both devices, while challenges in view include the lack of ground-truth data in most cyberattacks. Future research shall be on few-shot learning, lightweight model implementation, deep learning cutting-edge methodologies, penetration testing, and the use of additional sensor and actuator data to enhance the anomaly detection system.

Using the IoT 2023 dataset as a thorough benchmark, the study of [6] tackles the issue of feature extraction from IoT data. The goal is to gain a better understanding of the dataset's properties and possible uses by evaluating both classic statistical approaches and ML-based methods. Feature extraction takes on more significance in the context of the Internet of Things (IoT) because of the "curse of dimensionality," the well-known fact that data processing and analysis get more complicated as the number of dimensions grows. Various techniques have been surveyed to place them in the context of their strengths in capturing relevant information, reducing dimensionality, and improving performance in IoT analytics. Some key findings in this respect include the Hughes phenomenon: classifier performance may get better with more features up to some optimal point before deteriorating. In this paper, via ample experiments and performance analysis, guides the choice of suitable feature extraction methods to be deployed for various IoT applications. This will, therefore, help in the practical development of IoT solutions in 2023 and beyond. Besides, according to the authors, little effect of reducing features on the model performance is up to an accuracy of 93.04% using Decision Trees and 93.05% using Random Forest models.

This paper reviews CNN for anomaly detection within the Internet of Things networks [7] and, thus, tries to evaluate the performance of dimensions CNN1D, CNN2D, and CNN3D in the presence of normal and anomalous network data. It shows the models' trustworthiness in detecting different cyber-attack types and maintaining the integrity of the IoT network traffic. Various datasets are used, including IoT Network Intrusion, Bot-IoT, MQTT-IoT-IDS2020, IoT-DS-2, IoT-23, and IoT-DS 1. This study concludes that CNN2D and CNN1D are very good at identifying anomalies in IoT networks because of how accurate and fast they are. Thus, from the current and future perspectives, these models are very promising in building a solid structure for intrusion detection in the network of computers. Moreover, the authors recommend that future research should be directed along the lines of other deep learning approaches, such as FFN, RNN, and GAN, that are suitable for transforming this system into a high anomaly detection one to rise to the challenges of the shifting paradigm in cyber security.

Using ML approaches, Prazeres et al. (2022) evaluates AI-based malware detection in IoT network traffic. The study makes use of the IoT-23 dataset using real IoT network traffic of both benign and malicious, including numerous forms of malware including botnets and DDoS attacks. Key strategies to categorize network traffic are feature selection, data normalization, and the application of several ML models (Logistic Regression, RF, ANN, and Naïve Bayes) [41].

The authors of [20] evaluate different algorithms of anomaly detection and classification using the IoT-23 dataset. They found that, out of those, the RF algorithm was most effective with an accuracy of 99.5% and precision. It can be seen that ANNs have biases toward classes with higher occurrences, possibly due to neuron weight configurations, and the Support Vector Machine turned in the poorest result at an accuracy of 60% which can't predict benign captures, but it turned in a relatively high recall rating. The study concludes that RF is the best algorithm for detecting and classifying anomalies in the IoT-23 dataset, which was also revealed in the past by related studies and proposes further research into the causes of high accuracy by simpler models and the potential of advanced neural networks to enable improved performance. Especially in multi-class issues, Random Forest has repeatedly shown to be a useful classifier in identifying intrusions. For example, with high accuracy rates in multi-class classification, a Random Forest application to the IoT23 dataset revealed better performance than other machine learning models [8].

Likewise, XGBoost, known for its boosting power, has demonstrated extraordinary intrusion detection capability. As network intrusion detection systems built for IoT networks show, XGBoost lowers mistakes and raises classification accuracy by iteratively strengthening weak models [9]. Research also shows how well it can handle prevalent network dataset class imbalance problems [10].

Intrusion detection systems have used ANNs to learn intricate connections inside network traffic data. Their versatility and great performance while managing big datasets help them to shine in multi-class detection challenges. In one study, for example, ANNs were tested against various supervised learning techniques; Random Forest turned up as the best performance for multi-class intrusion detection. Nevertheless, ANNs also displayed competitive performance in precisely categorizing fraudulent traffic [11].

CNNs' capacity to automatically learn geographic information from data has driven their increasing application to IDS. CNNs have proved quite effective in the framework of multi-class classification in spotting network breaches. For multi-class detection on the Bot-IoT dataset, for example, a CNN model was able to reach accuracy values near 99.9986% [12]. To increase the long-range dependent detection capability in network traffic, CNNs have also been coupled with other DL models such Recurrent Neural

Networks (RNNs). There has been a considerable improvement in the detection accuracy of learning the spatial and temporal components of the network traffic data when CNN is paired with Gated Recurrent Units (GRU) [13].

Using multiclassification models within the PySpark architecture, a 2024 study by Alrefaei et al. offers IoT network real-time intrusion detection system (IDS). One-Vs-Rest (OVR) method machine learning approaches include Random Forest, Decision Trees, Logistic Regression, and XGBoost help to enhance detection accuracy and minimize prediction latency. Class imbalance is solved via data cleansing, scaling, and SMote using IoT-23 dataset. Random Forest displayed the fastest prediction time at 0.0311 seconds, but XGBoost achieved the highest accuracy at 98.89%, so underlining the system's value in real-time IoT threat detection and so reducing security concerns[15].

Combining the advantages of several classifiers, ensemble approaches have shown notable gains in IDS performance. In particular, hybrid models, which combine classifiers in voting systems to improve overall accuracy and detection rate, have encouraging results. Upadhyay et al. (2021) presented a majority voting ensemble combining Random Forest, XGBoost, and KNN with additional classifiers for SCADA-based power grids. Their model demonstrated gains in binary and multi-class classification by selecting features using Recursive Feature Elimination and then utilizing majority voting to increase precision and recall [1].

Similarly, Hussein et al. (2021) achieved greater detection rates across several assault categories by integrating AdaBoost, Random Forest, and XGBoost in an ensemble strategy for multi-class and binary classification. Their approach was verified on datasets like NSL-KDD and UNSW-NB15, and the ensemble performed better than individual classifiers regarding accuracy and precision [14].

A hybrid model that included Random Forest, XGBoost, and KNN and was improved using feature selection approaches was also investigated by Liu et al. (2023). When tested on many datasets, their approach showed an increase in overall detection accuracy and a decrease in false positives. In this study, the voting system of a hybrid classifier, which combines other individual classifiers, is achieving better performance than stand-alone classifiers [16].

Voting classifiers have emerged as a popular IDS technique, especially in IoT contexts where various threat types necessitate adaptable detection systems. A voting classifier that included AdaBoost, Random Forest, KNN, and SVC was suggested in research by Mhawi et al. (2022) to handle the high-dimensional nature of network traffic. Particularly in situations involving multi-class classification, their hybrid model outperformed solo classifiers. On the CICIDS2017 dataset, the model showed accuracy gains of up to 99.7% [17].

Research shows that a voting ensemble of classifiers including RF, XGBoost, AdaBoost, KNN, and SVC performs better than any classifier taken on alone. Leevy et al. (2021) evaluated different ensemble models, including XGBoost and Random Forest, to find assaults in the framework of the Internet of Things. Their studies show that in terms of adaptation and accuracy, ensemble models usually outperform individual classifiers [18]. Rahman et al. (2021), who combined AdaBoost and XGBoost with different feature selection methods, further highlight the significance of feature selection in these hybrid models. Their hybrid model decreased computing overhead, which is crucial for real-time intrusion detection systems while increasing detection accuracy [21].

When testing intrusion detection systems (IDS), the gold standard is the IoT23 dataset, which simulates real-world IoT network traffic. It provides several attack scenarios that researchers can use to test the multi-class classification capabilities of DL and ML models. Detection rates surpass 99% when models based on Random Forest and CNN are used [19].

# CHAPTER 3

# DATA PRE-PROCESSING

Pre-processing in the context of data science is the set of steps used to prepare raw data for machine learning models to examine. This stage is crucial since, unbridled, the irregularities, errors, and incompleteness of real-world data could lead to poor model performance. For the purpose of facilitating the model's learning, pre-processing primarily aims to clean, convert, and organize the data. The following are some of the actions done in order to prepare the data:

## 3.1 Data Analyzing

Data analyzing is a crucial part of data pre-processing. This important stage helps one to figure out the type of dataset, class distributions, patterns and some other valuable information which will help one to plan for how these data needs to be prepared. In this work we are using IoT23 dataset. Below we will talk about the class distribution of this dataset for both standalone and hybrid models.

### 3.1.1 Class Distribution for Standalone and Hybrid Models

For the standalone ML and DL models along with the hybrid models we had used 100,000 data for binary classification and for multi class classification we had used 69,398 data. For binary classification each class has 50,000 observations and for multi class classification each class has around 10,000 observations. In binary classification there are total of 2 classes which are Benign and Malicious and for multi class classification there are total 7 classes which are Benign, C&C- HeartBeat, DDoS,

Okiru, PartOfHorizontalPortscan, C&C, and Attack. The visualizations of these classes are given below:

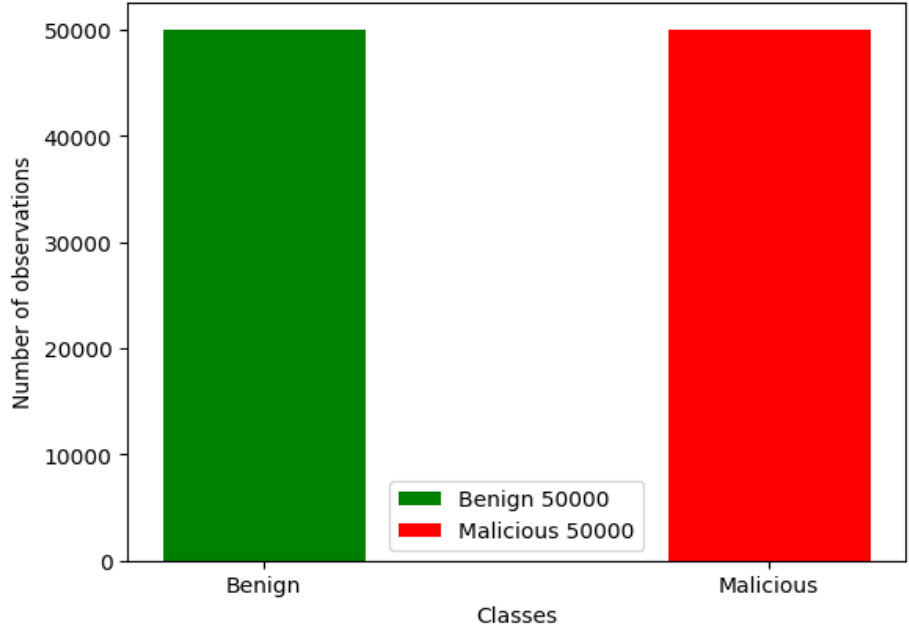

Fig 1: Class Distribution for Binary Classification

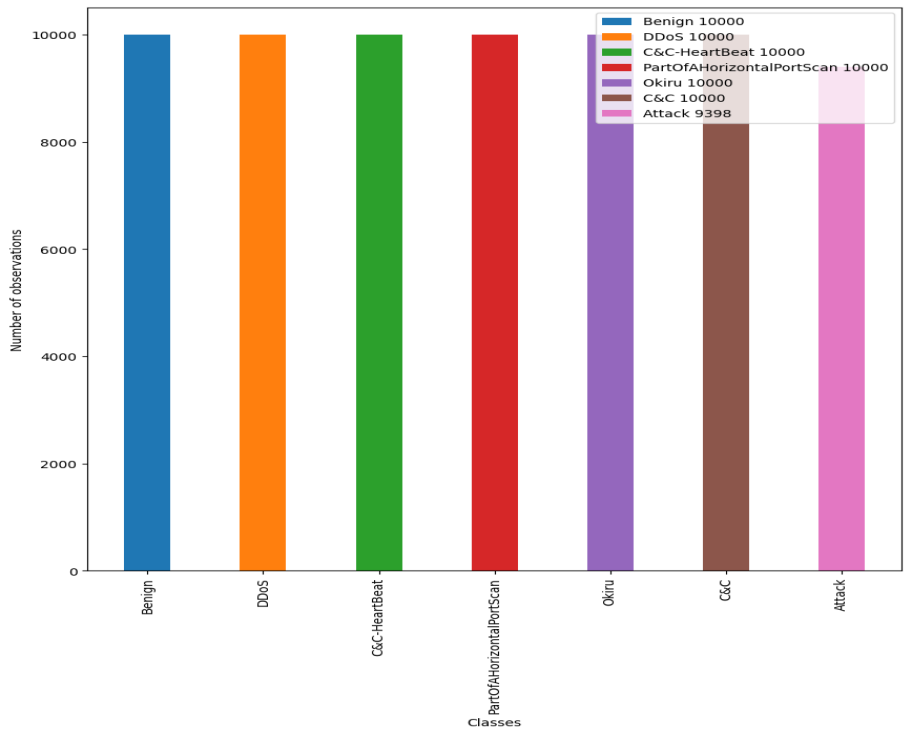

Fig 2: Class Distribution for Multi Class Classification

## 3.2 Missing Data Handling

Missing data handling refers to how to handle missing values such that they don't negatively impact target prediction. In order to address the missing values in the IoT23 dataset, we substituted 0 for the empty values of numerical features. Additionally, there was a categorical feature called service that had been filled with the value "unknown" and also had empty values.

## 3.3 Categorical Feature Conversion

To train the machine learning models we must need to convert the categorical features into numeric values because the models cannot understand word as an input value so, to do that we had applied a method called one hot encoding and this method converts the features like if a specific class is present in an observation then it denotes the value of that observation for that class as 1 otherwise 0.

## 3.4 Feature Engineering

The goal of feature engineering in machine learning is to improve a model's prediction power by enhancing it with additional features or by tweaking its current features. It is the conversion of raw data into a format better suited for model training so the computer may more effectively find important trends, correlations, and patterns.

For my dataset, first we extracted information from two of the features containing the IP addresses. The first information we extracted tells us if the IP is private or global and the second information tells us the countries of the IP addresses. And after that we removed those IP Address columns because it will overfit the model with training data which is not good for our model. There are some more id columns that are also being removed. After removing the id columns Permutation Importance test was done on all the remaining features and from the test results it showed that removing history will not impact on predicting the target because if we remove that then the decrease in the accuracy score is negative or unchanged so after this test, we have removed those that features also. Below is the visualization of the test:

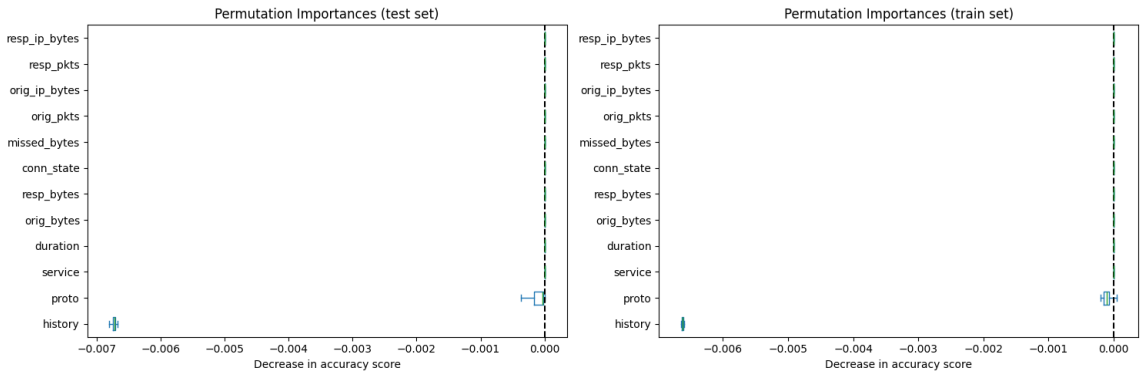

Fig 3: Permutation Importance Test

After doing the feature engineering we had a total of 36 features.

**3.5 Data Splitting**

In machine learning, data splitting is the technique of partitioning a dataset into several subsets for model training and evaluation. This stage is crucial to guarantee that, instead of only memorizing the training data, a phenomenon called overfitting a machine learning model generalizes successfully to new, unknown data. Below we will discuss the data splitting ratio for both standalone and hybrid models.

**3.5.1 Data Splitting for Standalone Models:**

For standalone models, we had segregated the dataset into a train, test and validation segment. We had used 70% for training, 20% for testing and 10% for validation. Below is the bar graph of that distribution:

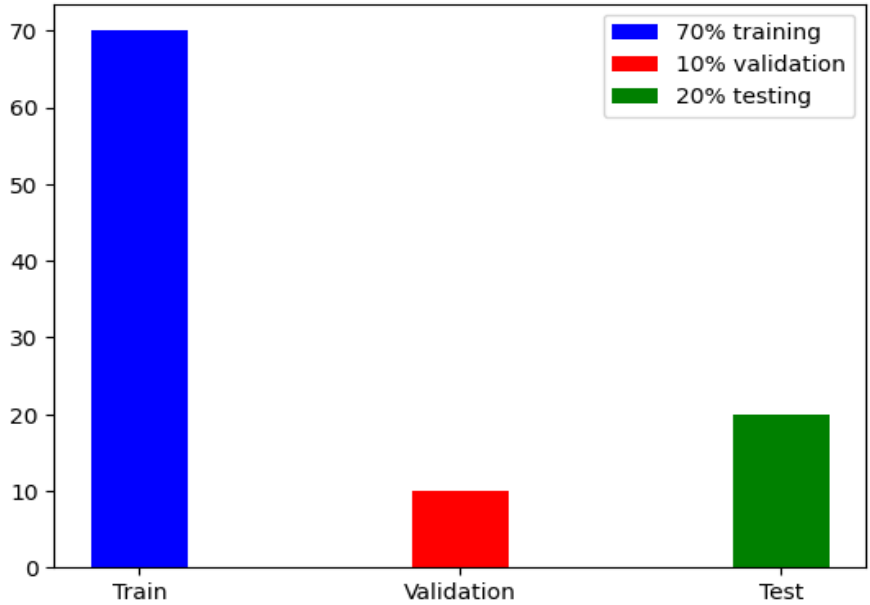

Fig 4: Data Splitting for Standalone Models

**3.5.2 Data Splitting for Hybrid Models:**

For hybrid models, we had taken 80% of the total data for training and 20% data of the total data for testing and as we are using cross validation which we will talk about later, it will randomly split the training dataset into validation and training data. Below is the visualization of data splitting:

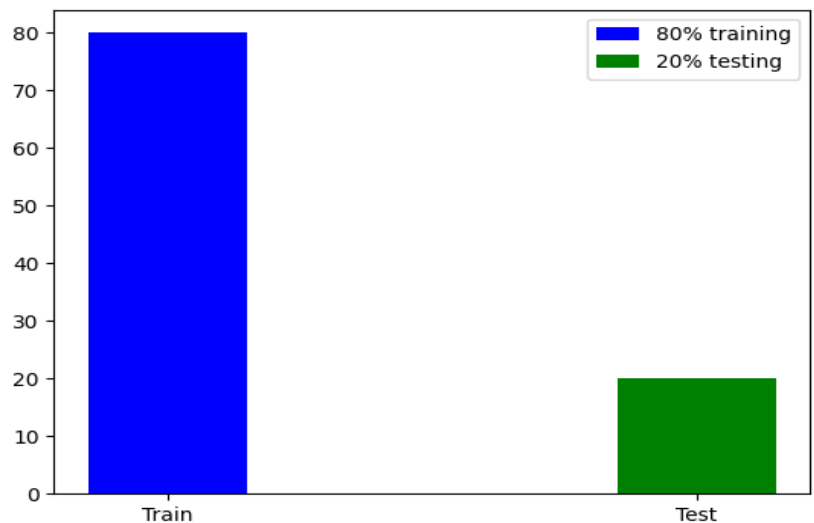

Fig 5: Data Splitting for Hybrid Models

Now we will talk about K-Fold Cross Validation.

**3.5.2.1 K-Fold Cross Validation**

K-Fold Cross Validation is a resampling method for assessing ML models that involves splitting the data into k equal-sized folds. In each iteration, the validation set is one-fold, while the training set consists of the remaining K-1 folds. A more accurate evaluation of the model's efficacy can be obtained by averaging the results of this k-times method. Although it aids in avoiding overfitting, K-fold cross validation offers a more reasonable assessment of a model's ability to generalize.

For our models we had used 5-fold cross validations which splits the training set 5 times into 5 different training and validation datasets.

**3.6 Feature Scaling**

In machine learning, feature scaling is a preprocessing method used to change the range or distribution of data values thereby guaranteeing consistency across features. For algorithms sensitive to changes in feature magnitude, such SVM) and KNN, this stage is essential since it standardizes or normalizes the data to a uniform scale, hence lowering the variability among features.

Eliminating bias resulting from features of different magnitudes which could skew model predictions are dependent on feature scaling. By lowering variations in feature range, techniques such as normalization and standardization help machine learning models to operate better [22].

For my dataset we used Min-Max Scaler and let's talk about it now.

### 3.6.1 Min-Max Scaling

Min-Max in ML, scaling is a normalizing technique used to rescale traits within a certain range—usually [0, 1]. This approach guarantees that no feature dominates the others and is particularly useful in cases of variable scales or units for the features. The conversion is accomplished with the following formula:

$$X' = \frac{X - X_{min}}{X_{max} - X_{min}}$$

Where $X'$ is the value after scaling, $X$ is the original value of that particular observation, $X_{min}$ is the minimum feature value, and $X_{max}$ is the maximum feature value.

### 3.7 Challenges

During pre-processing the data, we have faced 3 crucial challenges which are huge data volume, class imbalance and information leakage and now let's talk about them in brief.

### 3.7.1 Data Volume

Data volume plays an important role in machine learning. For our IoT23 dataset we had 21GB of data with around 60 million observations which is making the pre-processing tasks like feature scaling, one hot encoding, and label encoding very slow. Moreover, it was making the training process very slow too and that's why to mitigate these problems we had taken a portion of data for binary classification, multi-class classification and hybrid approach which will not affect the model performance in a bad way and also after doing that, the pre-processing tasks became very fast along with the training process.

### 3.7.2 Class Imbalance

In ML, class imbalance is the result of an unbalanced distribution of classes in a dataset whereby one class (or a small number of classes) has significantly more instances than other classes. This usually happens in classification problems when some events or results are rare in relation to others. There was a class imbalance in our dataset consisting of eleven assault kinds or classes; as a result, we removed the classes with the fewest observations and kept seven that had a sufficient amount of data.

### 3.7.3 Information Leakage

In machine learning, information leakage is the result of a model inadvertently accessing unrestricted material from the training set, hence producing poor generalization and unrealistic performance measures and possibly undermining the predicting capability of the model. In our case after the preprocessing when we started training the models we have seen that from the very first epoch the accuracy is too high like over 90% which is not sophisticated because in the very first epoch there is no model which can be perfect or close to perfect so after getting this kind of output, we had suspected that there is a information leakage which was during the data scaling. Before knowing the problem we were scaling the whole dataset altogether and at that time the scaler we had used gained access to all the data which was not supposed to be happened and scale the data according to that but after finding the problem at first we split the data and then we fit our scaler on the training data then we used that scaler to transform our training, testing and validation dataset and in this way the scaler was not gaining access to the validation and testing data.

# CHAPTER 4

# BINARY CLASSIFICATION USING STANDALONE MODELS

## 4.1 Chapter Background

In this section we will talk about ML and DL algorithms along with the tools that had been used for the binary classification. So far, we have used 2 different machine learning models which are RF and XGBoost and for the deep learning models we had used ANN and CNN. For the programming language we used python because it is easy to use, and the libraries are too rich for this kind of work. For the libraries we had used Scikit Learn, Keras, and TensorFlow to build and train the models.

In the next sections we will talk about the methods in detail, how they work and the results.

## 4.2 Utilizing Machine Learning Techniques for IoT Security

### 4.2.1 Methodologies

Various machine learning techniques are used in the framework of IoT security to improve anomaly detection and threat mitigating:

**4.2.1.1 Artificial Neural Network (ANN)**

ANNs are computational models that mimic the way the human brain works; they enable the discovery of intricate patterns in massive datasets. ANNs perform well in IoT due to their inherent learning and adaptive capabilities in dynamic environments that make them for detecting sophisticated cyber threats.

**4.2.1.2 Convolutional Neural Network (CNN)**

IoT applications that examine time series and sensor data have embraced convolutional neural networks (CNNs), despite CNN being traditionally used for location and picture data. With their excellence in feature extraction and hierarchic learning, they can recognize complex patterns within streams of IoT data and thus improve anomaly detection capability. All of these algorithms offer robust tools to secure IoT networks against a wide array of cyber threats.

**4.2.1.3 Random Forest**

As part of the ensemble learning approach, Random Forest (RF) generates predictions for classification problems by use of a large number of decision trees that are trained together. RF benefits IoT security through enhanced predictive accuracy and reduced over-fitting of intrusion detection systems, thus making them more reliable.

**4.2.1.4 Extreme Gradient Boosting (XGBoost)**

XGBoost is a most refined ensemble method designed to couple the predictions of several weak models into a strong model. With the added ability to process large-scale and sparse data, it can very well be applied in IoT environments characterized by heterogeneous voluminous data.

### 4.2.2 Application Process of Machine Learning to IoT

Application of machine learning in IoT ecosystems requires design considerations that take into account the unique characteristics and challenges of an IoT environment. The first consideration is related to data diversity and heterogeneity: IoT devices generate massive amounts of data which differ considerably by type, format, and quality. This calls for efficient pre-processing techniques that ensure consistency and reliability in this diverse data. Proper techniques—normalization of the data, noise reduction, handling missing values—are in order during the preparation of data for ML algorithms. It factors IoT devices' "resource constraints." The most IoT devices have limited energy, memory, and processing capability. Consequently, lightweight ML models include Decision Trees and improved variants of more intricate techniques like Extreme Gradient Boosting are more sought for. These models should be built and applied in a way that reduces the consumption of resources as low as feasible while preserving great accuracy and performance. "Scalability" is another essential consideration. This is so because IoT networks can comprise hundreds or thousands of devices, each generating continuous data streams. For this reason, ML algorithms need to be scalable to handle this huge amount of data efficiently. Distributed computing and edge computing, where data processing happens closer to the source of data, may decrease problems of scalability by reducing latency and bandwidth usage for sending data to some central servers. Real-time processing is essential for effective security in IoT devices. Given the dynamic nature of the environment in an IoT, to function effectively, the ML models have to be developed with real-time data analysis and threat detection capabilities. Algorithms such as CNN and ANN can be optimized for real-time inference that identifies anomalies and a reaction to intrusion promptly. Moreover, it is capable of using online learning methods so that it will keep updating the model whenever new data arrives to make it adapt to changes in patterns or newly emerging threats. This will, therefore, make "robustness and resilience" the major priority in the face of adversarial attacks. Sophisticated cyberattacks against IoT devices constantly try to dupe these ML models. Techniques such as adversarial training enhance the robustness of ML models by training them using normal examples and adversarial examples. Techniques like RF can add another line of defense by fusing several models to increase detection accuracy and hence minimize successful attacks. Lastly, there are critical considerations of "privacy and security of data". In many cases, IoT devices gather sensitive information, thus making the question of the privacy of data a serious one. It is possible to enhance privacy and security by training models using privacy-preserving machine learning techniques such as federated learning on decentralized data without having to move the raw data. In addition, ensuring compliance with data protection regulations and applying encryption procedures at all levels offers another layer for guaranteeing the integrity and confidentiality of data. In other words, any effective application of ML to IoT would require a multi-dimensional approach to issues of data heterogeneity, resource-constrained devices, scalability, real time processing, robustness to adversarial attacks, and data privacy. Provided that these challenges are taken into account, the design of effective and efficient ML solutions can be achieved through tailoring to meet the unique demands of IoT environments for bolstering security and functionality in IoT networks.

## 4.3 Evaluation of Proposed Methodologies

### 4.3.1 Evaluation Analysis Methods

We used a set of well-known criteria to assess the performance of the above-described algorithms; they will be covered in great detail in the "Results" part. One should explain four basic ideas before exploring these measures [23]:

- **True Positive (TP):** The total count of real positive cases the model accurately found.
- **True Negative (TN):** The total number of actual negative cases the model accurately found.
- **False Positive (FP):** The total number of actual negative events wrongly labelled as positive by the model.
- **False Negative (FN):** The total count of real positive cases mistakenly labelled as negative by the model.

#### 4.3.1.1 Precision

Calculating the fraction of correctly detected positive examples helps one to determine the performance of a model by means of a precision score. It is stated as TP to the sum of TP and FP ratio.

#### 4.3.1.2 Recall

Calculating the fraction of real positive cases that were accurately detected helps one assess a model's performance using recall score. It is defined as TPs to sum of TPs and FN ratio.

### 4.3.1.3 Accuracy

One way to evaluate a model's efficacy is to calculate the proportion of correct predictions relative to the total number of forecasts. It is the sum of all the cases, TP, TN, FP, and FN combined, divided by the total number of cases.

### 4.3.1.4 F-1 Score

Combining recall with precision into a single measure, the F1 score is a metric used to assess model performance. It is described as a harmonic average of recall and precision, therefore balancing the two measurements.

**4.3.2 Evaluation Results and Summary**

Within the light of the evaluation results presented in Table 1, our model performance was compared to the one reported in a referenced study [5], denoted as paper-1, which we observed an improvement in all metrics assessed. Its F1-score, recall, accuracy, and precision in the DNN(ANN) were 0.997, 0.997, 0.998, and 0.998, respectively; in comparison, paper-1 values were 0.940, 0.920, 0.930, and 0.970. That is a significant improvement of the model in detecting and classifying the anomalies occurring in the IoT networks. The comparison table is shown below:

| Model Name | Papers | Testing Accuracy | Precision | Recall | F1 Score |
|---|---|---|---|---|---|
| Random Forest | Ours<br>Paper 1 | 99.8<br>95 | 99.8<br>59 | 99.7<br>44 | 99.8<br>50 |
| XG Boost | Ours<br>Paper 1 | 98.9<br>N/A | 98.5<br>N/A | 98.7<br>N/A | 98.9<br>N/A |
| ANN | Ours<br>Paper 1 | 99.8<br>93 | 99.8<br>97 | 99.7<br>92 | 99.7<br>94 |
| CNN | Ours<br>Paper 1 | 94.0<br>N/A | 99<br>N/A | 88.92<br>N/A | 93.7<br>N/A |

Table 1: Model Comparison Table for Binary Classification

We also implemented the algorithm of RF, which gave very high scores of 0.998, 0.998, 0.997, and 0.998 in terms of accuracy, precision, recall, and F1-score, which were near perfect. This is very much different from inconsistencies seen in the paper-1 result, which reported 0.950 in accuracy and notably, the precision, recall, and F1-score values of 0.59, 0.44, and 0.50, respectively. Especially for the ANN(DNN) and RF methods, the methodologies described in Feature Engineering helped us to achieve significantly improved results when compared to paper-1. Paper-1 did not implement XG-Boost and CNN, as opposed to our work. Our XG-Boost model resulted in an overall accuracy of 0.989, along with very promising precision, recall, and F1-scores of 0.985, 0.987, and 0.989, respectively. On the other hand, the CNN model resulted in better results, in an overall accuracy of 0.940, with precision, recall, and the harmonized F1-score around perfect, i.e., 0.990, 0.889, and 0.937, respectively.

Below we will show some visualizations of Accuracy, Precision, Recall and F1-Score:

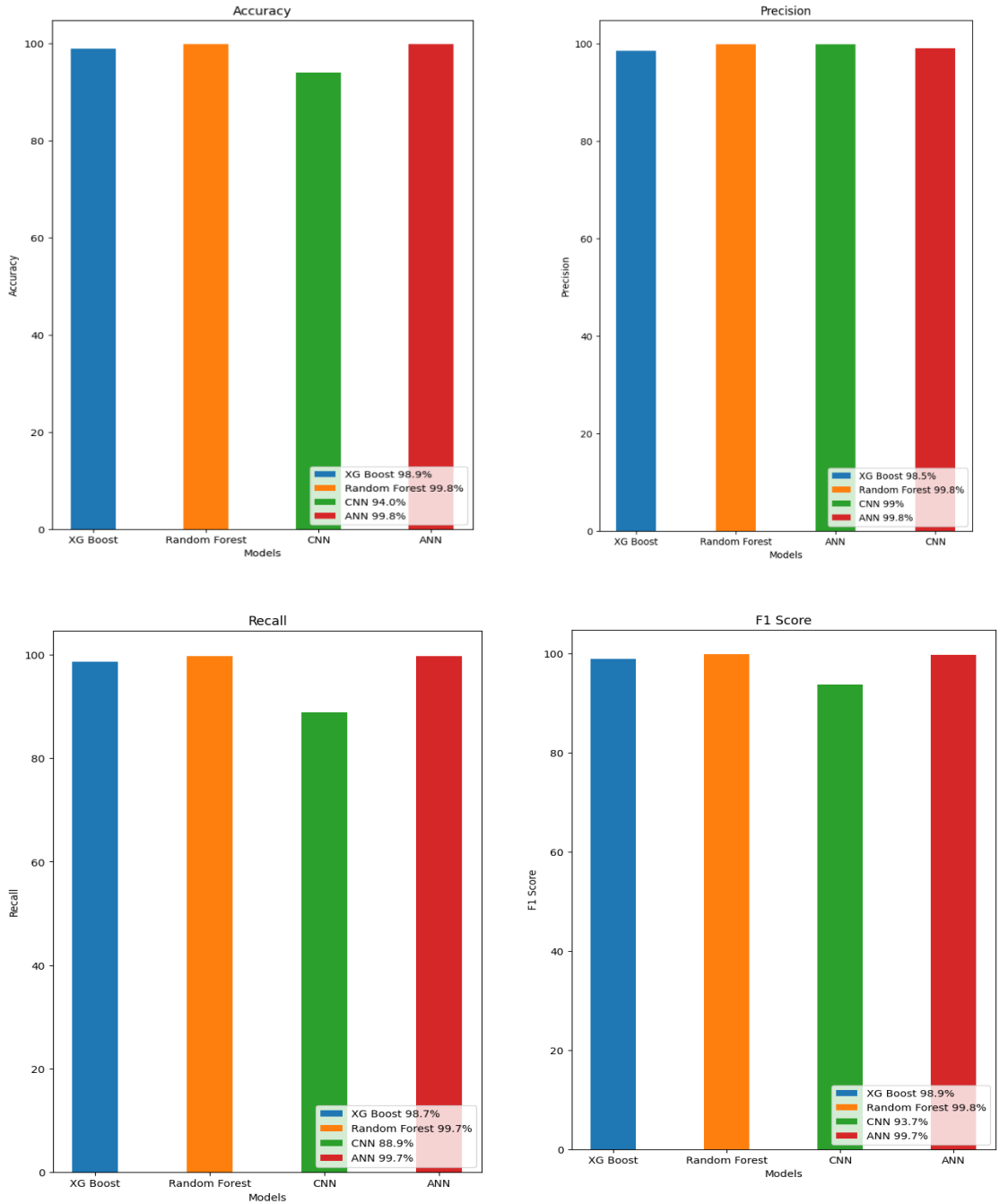

Fig 6: Bar Graph of Accuracy, Precision and Recall for Binary Classification

Now let's talk about the loss curves of these models.

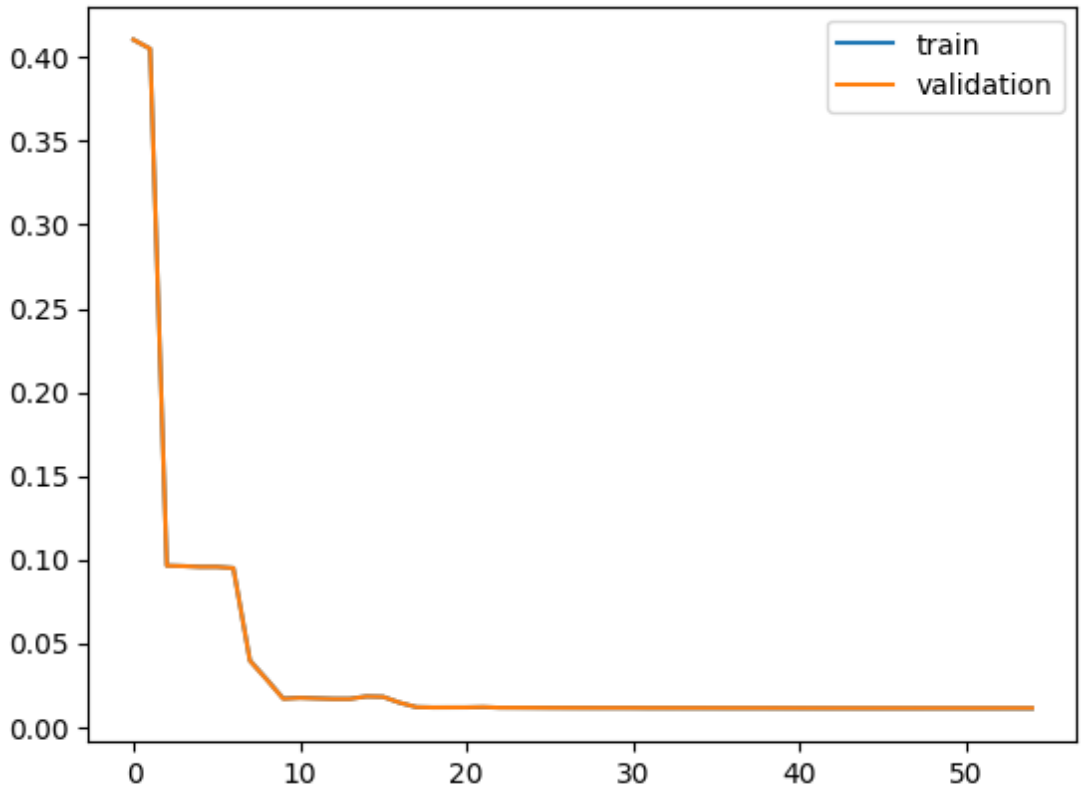

Fig 7: Loss Curve of XGBoost Model

From the above loss curve of XGBoost model, we are seeing that at the beginning of the training the loss starts from 0.40 and then gradually it decreases to the global minima, which is around 0. However, there is no spike in graph, so it tells that there is no overfitting in the model, and it is converging smoothly. The figure also shows that after around 55 epochs the model finished training.

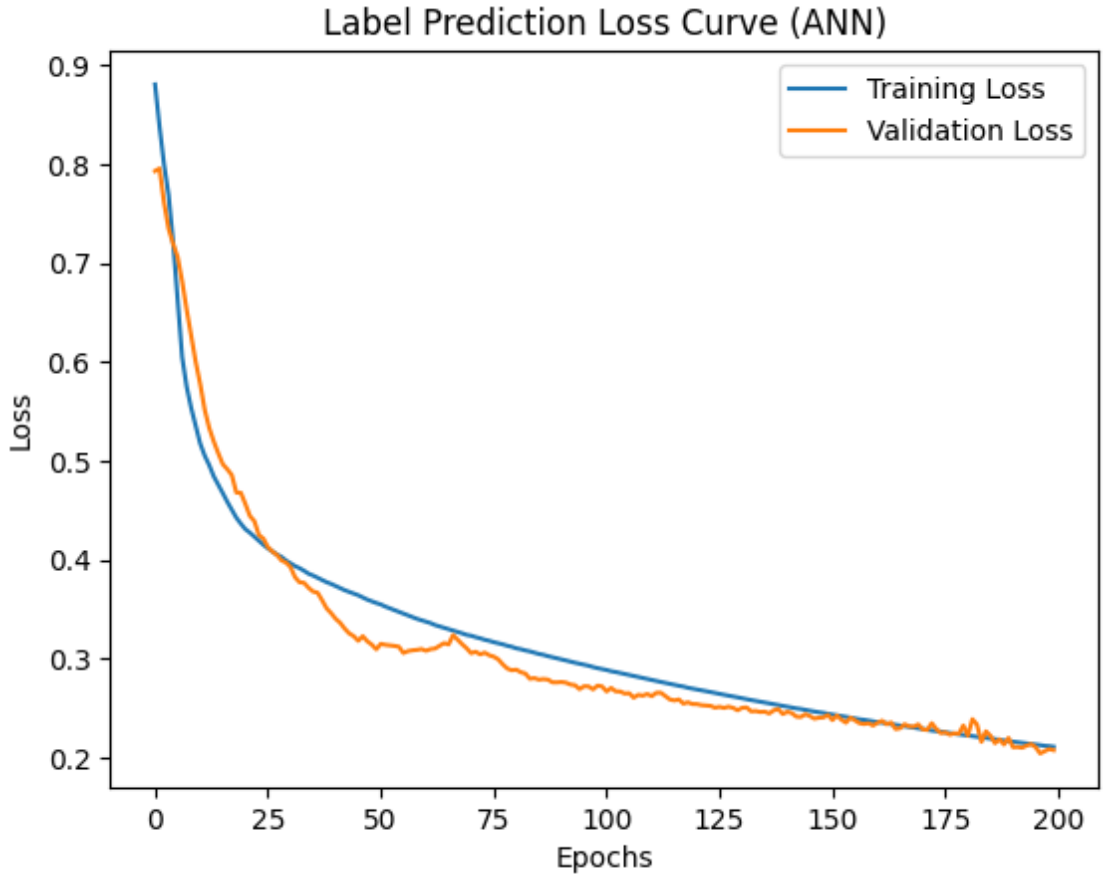

Fig 8: Loss curve of Artificial Neural Network (ANN)

From the above curve we are seeing that the loss of validation loss starts from 0.9 and gradually decreases to the global minima of around 0.2 at the end of training. We are also seeing that there are no spikes in the validation or the training curve, so both are converging smoothly. As the curves are converging smoothly there is no overfitting in the model.

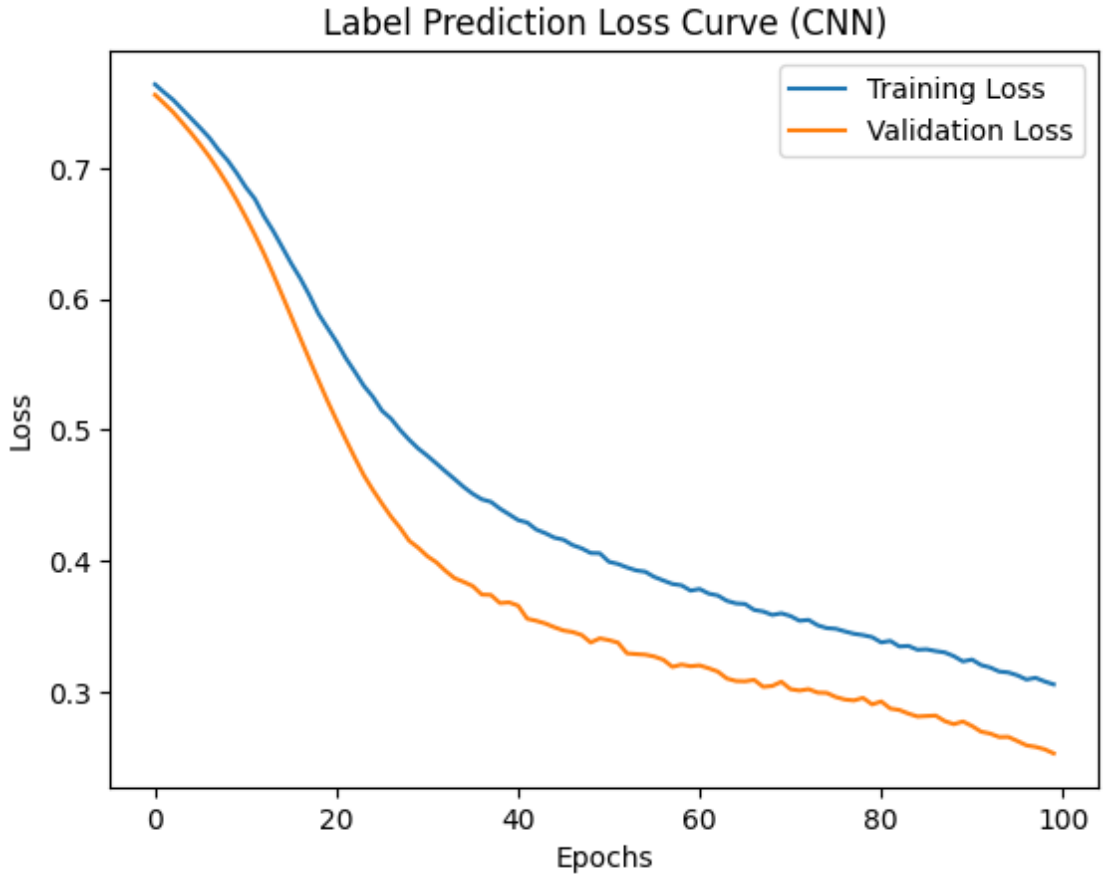

Fig 9: Loss curve of Convolutional Neural Network (CNN)

The above loss curve of CNN model shows that the training loss starts from around 0.75 and gradually decreases to around 0.35. On the other hand, for the validation curve it starts from around 0.74 and decreases to around 0.25. We are also seeing that there is no spike in the curves so both curves are converging smoothly. Moreover, there is no point where the validation curve crosses the training curve so there is no model overfitting.

## 4.4 Chapter Conclusion

After analyzing the bar graphs, loss curves and comparison table it can be said that RF, XGBoost and ANN are achieving very high score in Recall, Precision, Accuracy, and F1-Score which are better than CNN. But in terms of speed the XGBoost model is better than the Random Forest and ANN because it uses the gradient boosting technique to minimize the loss in each iteration, but RF takes the average of every tree output which can be very slow if you have bigger number of trees. On the other hand, for ANN if you have so many hidden layers or your dataset is very complex then it will take much more time than XGBoost to train the model. So, in terms of speed, accuracy and scalability for large and complex tabular datasets, the XGBoost model performs better than most of the models.

# CHAPTER 5

# MULTI CLASS CLASSIFICATION USING STANDALONE MODELS

## 5.1 Background

In this part we will discuss the tools applied for binary classification together with ML and DL methods. We have thus applied two distinct ML models: RF and XGBoost and for the DL models ANN and CNN. Python is the programming language we have picked since the libraries are too abundant for this type of task and it is simple to operate. For the libraries we had consulted to create and train the models Scikit Learn, Keras, and TensorFlow.

The approaches will be discussed in great length in the future parts, together with their outcomes and workings.

## 5.2 Methodologies

We had proposed 4 different methods for this classification task which are RF, XGBoost, ANN and CNN. Below are the short descriptions of those methods.

### 5.2.1 Random Forest

RF is a versatile ML technique that can be used for both regression and classification. It is one of the algorithms available in ensembles. The mode of the classes used for classification tasks can be generated by training with a large number of decision trees.

### 5.2.2 Extreme Gradient Boosting (XGBoost)

XGBoost is a ML technique contained in the gradient boosting framework and a subclass of ensemble learning. Starting with decision trees as basic learners, it improves model generalization via regularization techniques. Clear evidence of its computational efficiency includes its efficient processing, insightful feature significance analysis, and seamless handling of missing data.

Before talking about the ANN and CNN model let's discuss the activation functions and other parameters which have been used in those models.

### 5.2.3 Activation Function

An activation function in machine learning transforms non-linearity in the output of a neuron therefore allowing neural networks to learn and approximatively generate intricate patterns in data. Since it governs the information flow throughout the network, the activation function applied affects the performance, convergence rate, and correctness of a model.

In my ANN and CNN model we had used 3 activation functions which are Rectified Linear Unit (ReLU), Exponential Linear Unit (ELU) and SoftMax. Below are some of the brief descriptions of those functions:

#### 5.2.3.1 Rectified Linear Unit (ReLU)

To provide non-linearity into neural networks, deep learning makes great use of the basic and efficient ReLU activation function. It is defined as:

$$ReLU(x) = \max(0, x)$$

Here $x$ is the input. If the input is positive, the ReLU function outputs it straight; else, it returns zero. ReLU is especially fit for deep networks since this feature lets it escape the vanishing gradient issue that influences other activation functions such Sigmoid and Tanh [24].

**5.2.3.2 Exponential Linear Unit (ELU)**

Common in machine learning for solving gradient vanishing and bias shifts in ANNs, the ELU activation function is applied here. It is defined as:

$$ELU(x) = \begin{cases} x & if\ x > 0, \\ \alpha(e^x - 1) & if\ x \leq 0 \end{cases}$$

Here, $\alpha$ is a hyperparameter that controls the saturation for negative inputs. This structure is efficient in preserving the flow of information across the network and increasing model learning capacity since ELU keeps non-zero gradients for negative inputs [25].

**5.2.3.3 SoftMax**

In machine learning, especially in multi-class classification problems, the SoftMax activation function is extensively applied since it converts neural network outputs into a probability distribution over several classes. By exponentiating each output and normalizing it with the sum of all exponentiated outputs, SoftMax generates probability for every class that sums to one. This feature allows us to perceive predictions as probabilities; hence it is perfect for output layers in classification systems [26]. The function is defined as:

$$SoftMax(z_i) = \frac{e^{z_i}}{\sum_{j=1}^{n} e^{z_j}}$$

Here $z_i$ is the logit for class $i$, and $n$ is the total number of classes.

Now we will talk about the other three parameters which are Kernel Regularizer and Kernel Initializer or Weight Initializer and Loss Function.

**5.2.4 Kernal Regularization**

A kernel regularization is a machine learning method used to help prevent overfitting in models by incorporating a regularizing term in the loss function punishing high weights in the kernel (weight matrix) of a neural network layer. Usually used are regularizing techniques including L1, L2 (Ridge), or both (Elastic Net). Keeping the kernel weights small helps the regularization term—which influences the total cost function—to encourage the model to learn simpler patterns, hence improving generalization on unknown input. A study by Jiang et al. 2020 said that, by limiting the impact of high-complexity kernels, regularization improves the stability and performance of models, especially in high-dimensional feature spaces [27].

In my CNN and ANN models, we had used Elastic Net.

**5.2.4.1 Elastic Net or L1 L2 Regularization**

Elastic Net regularization is the result of combining L1 and L2 regularization. It provides a more reliable method of managing overfitting in machine learning models by combining the advantages of L1 (Lasso) and L2 (Ridge) regularization approaches.

- To promote sparsity, **L1 regularization** applies a penalty that scales with the absolute value of the model's coefficients. By zeroing some of the coefficients, this essentially does feature selection.

- The coefficients are uniformly shrunk but not necessarily driven to zero by **L2 regularization**, which applies a penalty equal to the square of the coefficients.

The Elastic Net regularization combines L1 and L2 penalties in the loss function, therefore integrating these two techniques:

$$Elastic\ Net\ Loss = Loss + \lambda_1 \sum |w| + \lambda_2 \sum w^2$$

Where $\lambda_1$ and $\lambda_2$ are the regularization strengths for the L1 and L2 penalties, respectively, and $w$ represents the model's coefficients. This approach is effective for feature selection, especially when variables are highly collinear, making it a versatile method in predictive modeling [28].

**5.2.5 Kernel Initializer**

In machine learning, a kernel initializer is the process of initial weight (or kernel) setting for a neural network before training. The initial choice influences the rate of convergence and the likelihood of avoiding issues including the disappearing or extending gradient problem, so influencing the training process.

Although there are other kinds of kernel initializers, Glorot (Xavier) Initializer, He Initializer, Glorot Uniform Initializer etc. is one of the common ones. Glorot Uniform Initializer was the choice we made in our ANN and CNN models since it aids in the activation function's prevention of gradient vanishes.

Now we will go over the Glorot Uniform Initializer:

**5.2.5.1 Glorot Uniform Initializer**

By avoiding the disappearing or expanding gradient issues, the Glorot Uniform Initializer, also called the Xavier Uniform Initializer, is a weight initialization method intended to enhance deep neural network training. A neural network layer's weights are initialized by selecting samples from a uniform distribution that falls within the range:

$$W \sim U(a, b) = \left(-\frac{\sqrt{6}}{\sqrt{n_{in} + n_{out}}}, \frac{\sqrt{6}}{\sqrt{n_{in} + n_{out}}}\right)$$

Here $W$ represents the weight matrix, $n_{in}$ is the number of inputs, $n_{out}$ is the number of outputs, $U(a, b)$ is a uniform distribution between $a$ and $b$.

**5.2.6 Loss Function**

A loss function is an essential component of DL and ML since it measures the discrepancy between the target values and the model-generated projected values. It evaluates the performance of the model and provides a basis for training optimization. Minimizing the loss function will help the model to learn from its errors and progressively raise its accuracy. From classification to regression, loss functions are fundamental for many projects since they direct the learning process by means of parameter modifications in every training cycle [29]. There are different kinds of loss function which are Mean Squared Error (MSE), Mean Absolute Error (MAE), Categorical Cross Entropy etc. For my ANN and CNN models we had used Categorical Cross Entropy. we will discuss about this loss function below:

**5.2.6.1 Categorical Cross Entropy**

Categorical cross-entropy loss is one often used loss function for multi-class classification problems. Representing the real label as a one-hot encoded vector gauges the difference between the expected probability distribution of classes generated by a model and the actual distribution. The following is the formula for categorical cross-entropy loss:

$$L = -\sum_{i=1}^{c} y_i \log(p_i)$$

where $c$ is the number of classes, $y_i$ is the binary indicator (0 or 1) if class label $i$ is the correct classification, and $p_i$ is the predicted probability for class $i$. The negative log likelihood of the right class probabilities is known as categorical cross-entropy, and it improves learning results by imposing a bigger penalty for confident but incorrect predictions [30].

Now we will discuss the structures of ANN and CNN model.

### 5.2.7 Artificial Neural Network (ANN)

Motivated by the biological neural networks observed in the human brain, a computational model frequently referred to as an ANN Applied to challenging tasks including pattern recognition, regression, classification, and more, artificial intelligence (AI) and machine learning Every node in an artificial neural network (ANN) replicas the activity of real neurons by consuming data, processing it, and sending the outputs to the layer above. ANNs are made from layers of linked nodes, that is, neurons. The structure of our ANN model is shown below:

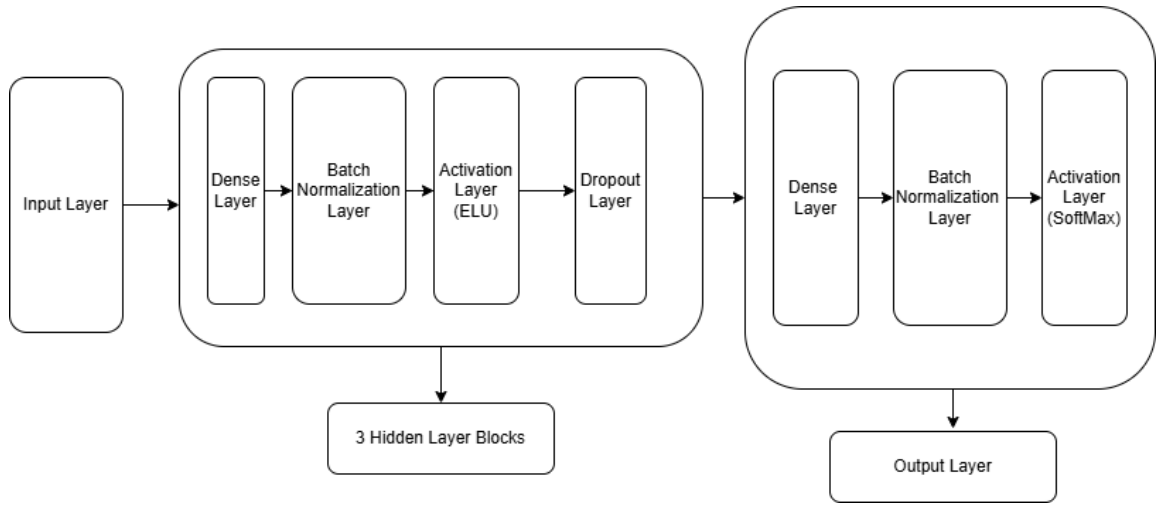

Fig 10: Artificial Neural Network Structure

The above architecture details the following: an input layer, three hidden layers, with each layer including a dense layer, a batch normalization layer, an activation layer using the ELU activation function, and a dropout layer. Furthermore, one dense layer, one batch normalization layer, and one activation layer with the SoftMax activation function make up the output layer of my artificial neural network model.

**5.2.8 Convolutional Neural Network (CNN)**

Designed largely for processing structured grid-like input, such images and videos, a CNN is a particular type of deep learning neural network. CNNs are well appropriate for jobs including picture classification, object detection, and facial recognition since they are rather good in spotting patterns and extracting hierarchical features from data. Our dataset is one-dimensional; hence we had employed 1D convolutional layers in the model even though my work is not image classification nor object recognition, we wanted to observe how the model works. The structure of my CNN model is shown below:

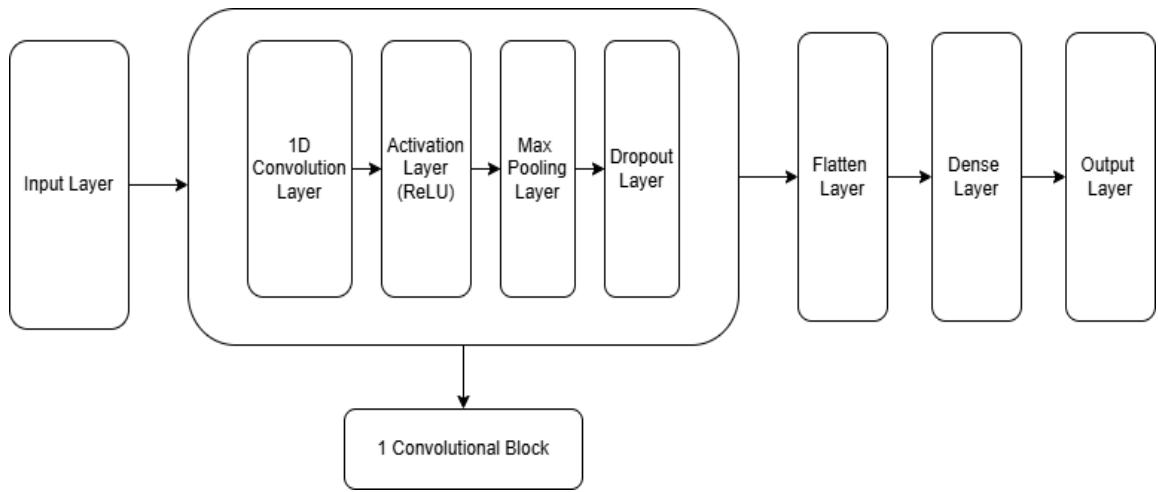

Fig 11: Convolutional Neural Network Structure

There is an input layer, 1 convolution block which is a combination of a convolutional layer, an activation layer with the ReLU activation function, a pooling layer with Max Pooling and a dropout layer. Furthermore, there is also one flatten layer, a hidden layer with the ReLU activation function, and an output layer with the SoftMax activation function after the convolutional blocks.

**5.3 Model Evaluation and Discussion**

Model evaluation is the process of assessing, on a given job, applying numerous criteria and techniques, the performance of a trained model. This is a crucial aspect defining the model's performance for the intended usage and degree of generalizing capacity to fresh data. Model assessment guides one to select the optimal model, adjust hyperparameters, and identify any problems including underfitting or overfitting.

We have evaluated our models based on 4 different scales: Precision, Testing Accuracy, F1 Score and Recall. Below are some brief descriptions of those metrics:

### 5.3.1 Accuracy

In ML or DL, accuracy is a performance metric used to show, in respect to all the predictions a classification model generates, how many right predictions it generates. Its definition consists as follows:

$$Accuracy = \frac{Number\ of\ Correct\ Predictions}{Total\ Number\ of\ Predictions} = \frac{TP + TN}{TP + TN + FP + FN}$$

Here,

    **TP (True Positive):** Correctly predicted positive observations

    **TN (True Negative):** Correctly predicted negative observations

    **FP (False Positive):** Incorrectly predicted positive observations

    **FN (False Negative):** Incorrectly predicted negative observations

Accuracy offers a clear evaluation of general correctness in multi-class and binary classification tasks; yet, it can be limited in imbalanced datasets, in which case alternate measures such as precision, recall, or F1-score could be more useful [31].

### 5.3.2 Precision

The proportion of actual positive predictions among all the predicted positive cases is known as precision. Here is the precision formula:

$$Precision = \frac{TP}{TP + FP}$$

Here,

    **TP (True Positive):** Correctly predicted positive observations

    **FP (False Positive):** Incorrectly predicted positive observations

When the cost of false positives is significant, precision is especially important since it shows how well the model can correctly identify pertinent events without over-predicting positives [32].

**5.3.3 Recall**

Recall also goes under the name sensitivity or true positive rate, a ML or DL statistic measuring a model's accuracy in spotting positive events. One definition of it is the ratio of genuine positive forecasts to the overall count of real positive cases or the sum of false negatives and true positives. The computation of recall follows this formula:

$$Recall = \frac{TP}{TP + FN}$$

Here,

**TP (True Positive):** Correctly predicted positive observations

**FN (False Negative):** Incorrectly predicted negative observations

It is crucial in fields like medical diagnosis or fraud detection, because overlooking pertinent cases (false negatives) could have major repercussions even if determining all important events depends on this [33].

**5.3.4 F1-Score**

One way to measure a model's efficacy in ML/DL is with the F1-score, which takes precision and recall into account. It provides a mix between the two metrics and shines in situations where class distributions aren't uniform or where false positives and negatives matter a lot. The F1-score computation equation is:

$$F1 - Score = 2 \times \frac{Precision \times Recall}{Precision + Recall}$$

Here,

**Precision:** Precision is the percentage of real positives to all expected positives.

**Recall:** The ratio of all actual positives to true positives is called recall.

When a balance between precision and recall is crucial, like in medical diagnostics or fraud detection chores, the F1-score is especially helpful [34]. Below are the graphs of average Accuracy, Precision, Recall and F1 Score of our four model:

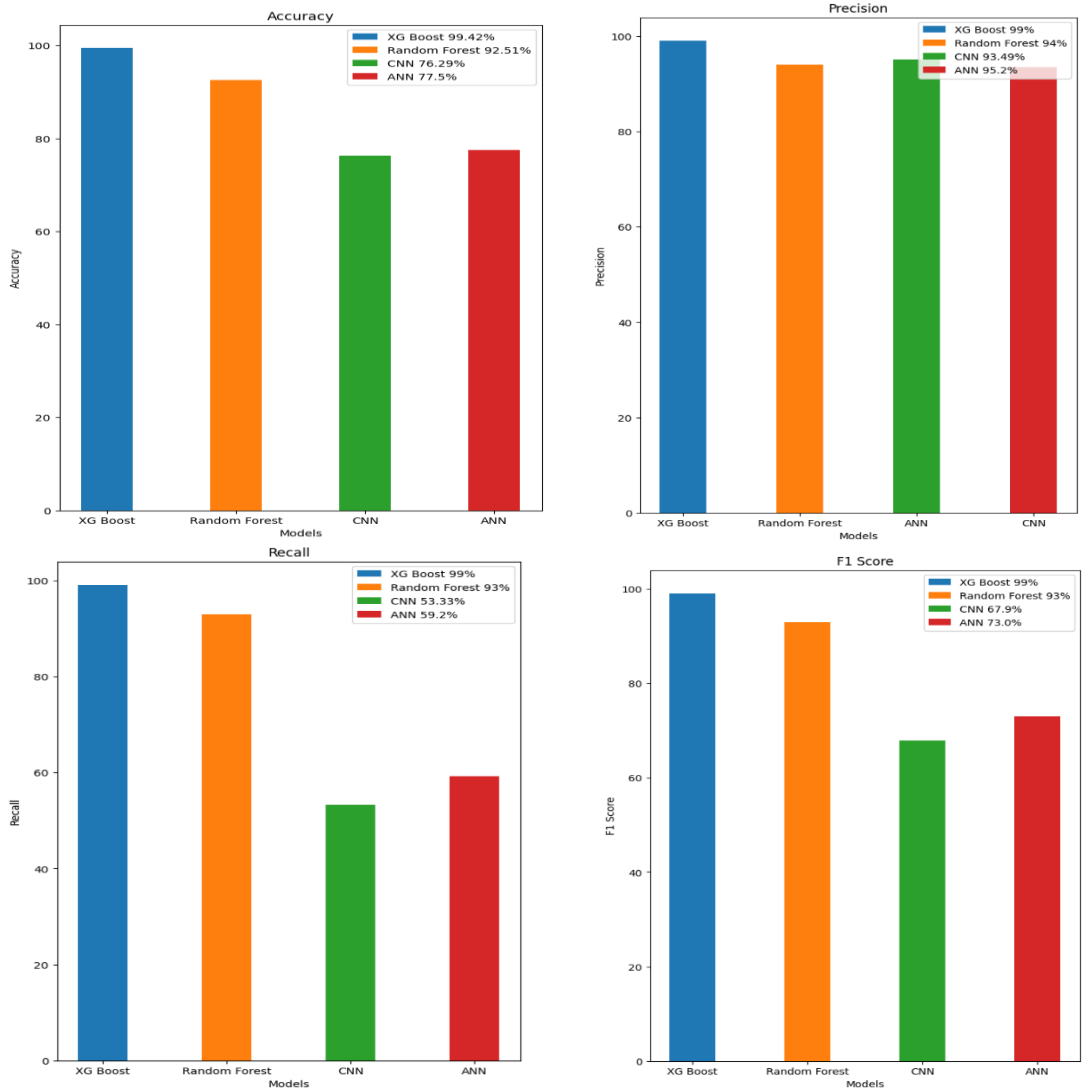

Fig 12: Bar Graph of Accuracy, Precision and Recall for Multi-Class Classification

From the above graphs we are seeing that XGBoost, Random Forest, CNN and ANN have 99.42%, 92.51%, 76.29% and 77.5% of average accuracy respectively. On the other hand, in precision bar graph XGBoost, Random Forest, CNN and ANN have 99%, 94%, 93.50% and 95.2% of average precision respectively. Furthermore, in the recall bar graph XGBoost, Random Forest, CNN and ANN have 99%, 93%, 53.33% and 59.2% of average recall respectively. In the F1 Score graph we are also seeing that XGBoost, Random Forest, CNN and ANN have 99%, 93%, 67.90% and 73% of average F1 Score respectively.

Now we will discuss the loss curves of the four models.

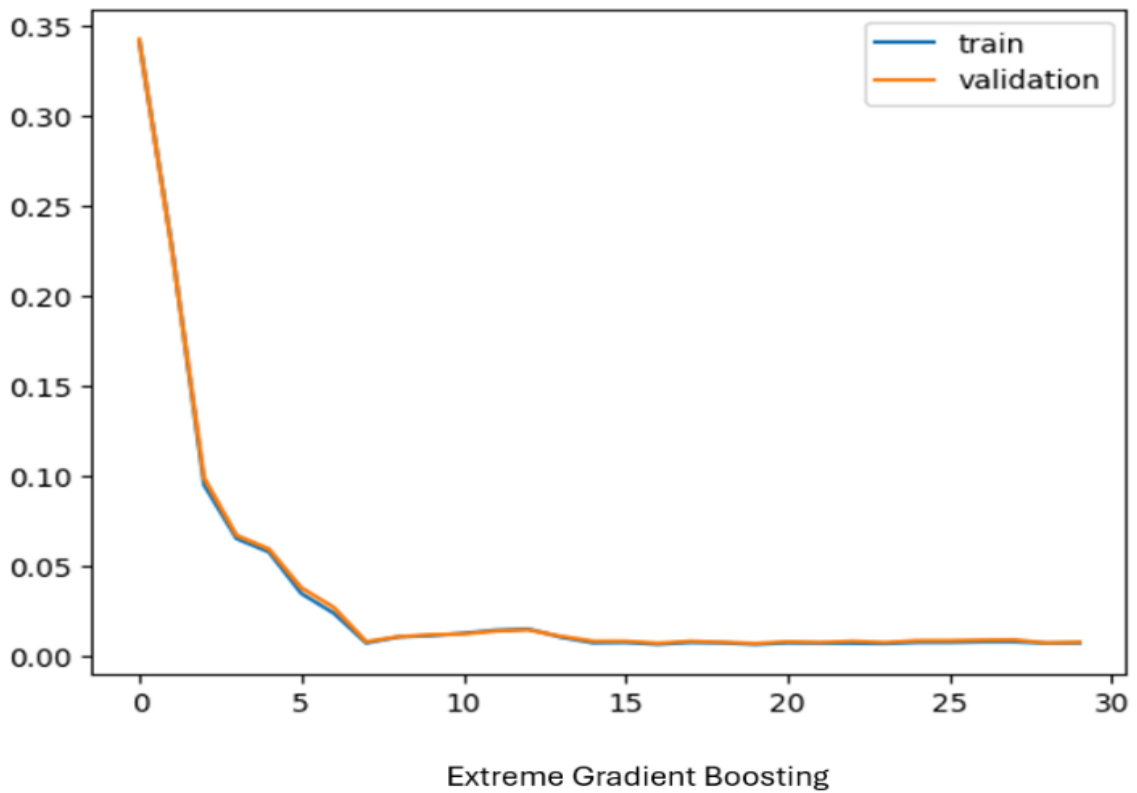

Fig 13: Loss curve of Extreme Gradient Boosting for Multi Class Classification

For the loss curve of Extreme Gradient Boosting (XGBoost) we are seeing that initially the loss was around 0.35 but gradually it has decreased and when the training end in 29th epoch the loss for both validation and training was close to 0 and we also see that as we are using early stopping it only took 29 epochs to finish the training. And both of the validation and training curves are smoothly converging with no fluctuation or spikes like there is no point when training curve is improving, and validation curve is not which tells us that there is no overfitting.

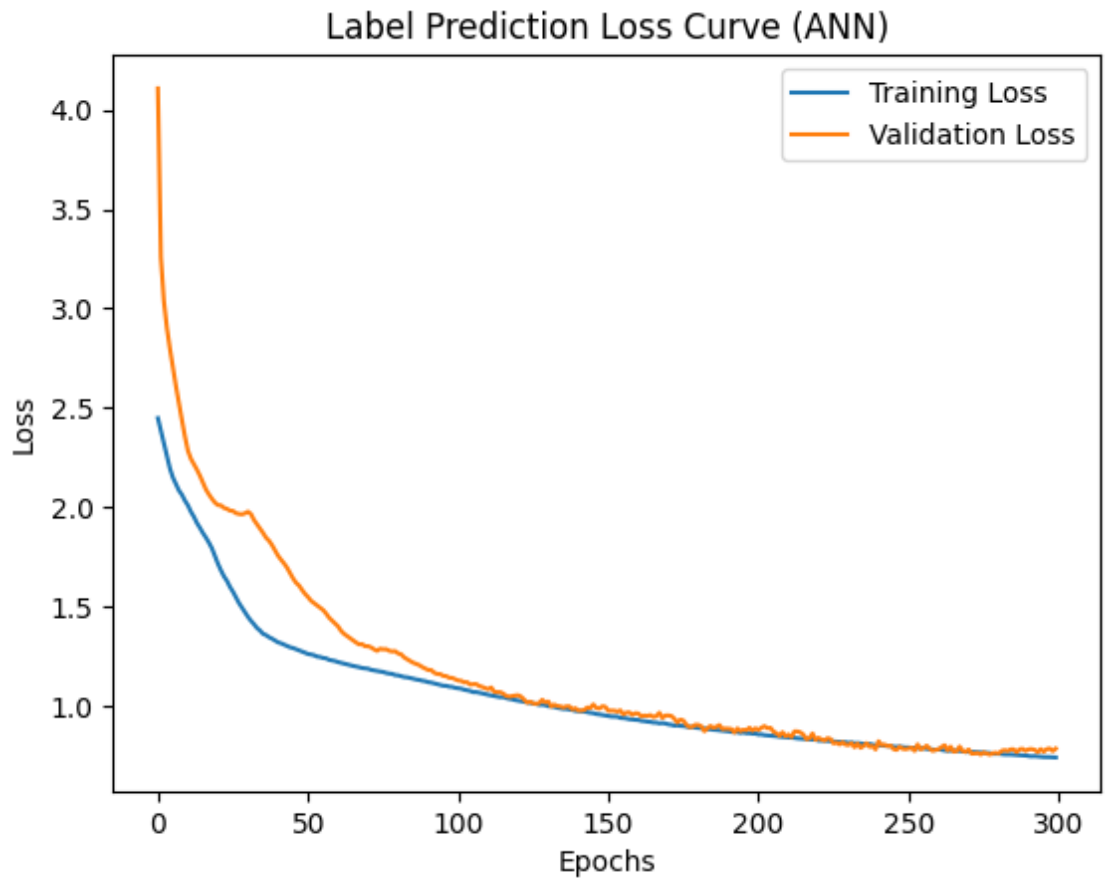

Fig 14: Loss curve of Artificial Neural Network for Multi Class Classification

For ANN the training and validation loss starts from between 4.0 to 4.5 and gradually decreases to its global minima which is around 0.5. We are also seeing that the curves are converging smoothly and also there is no sign of overfitting.

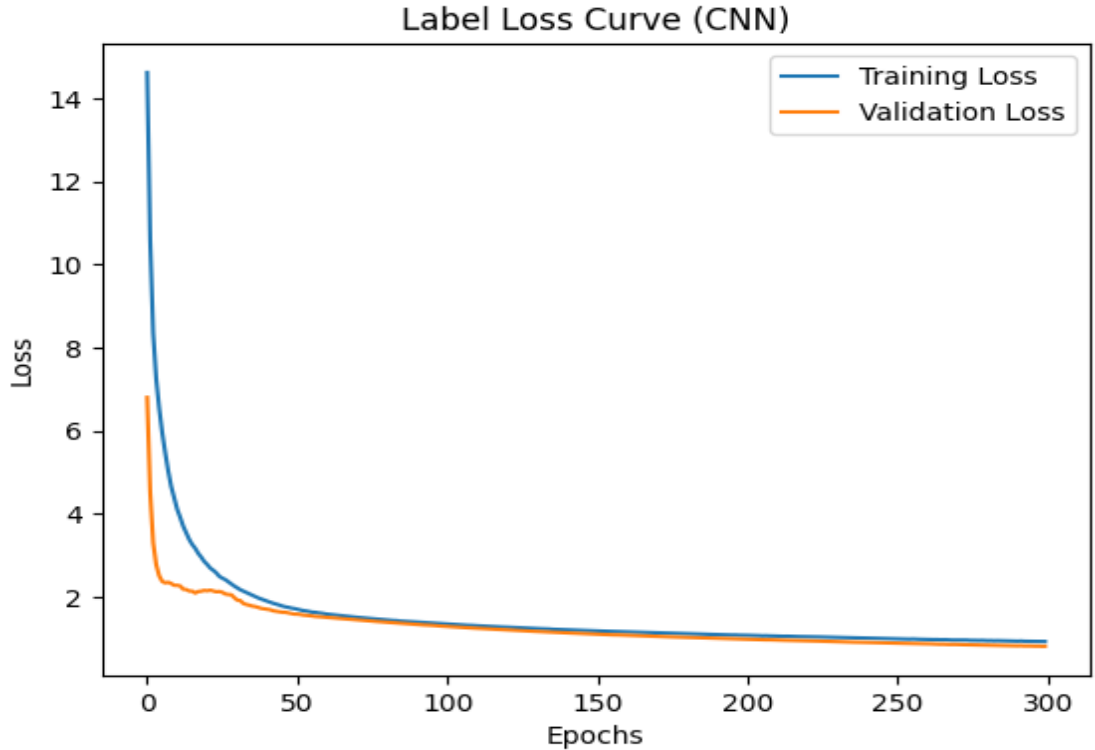

Fig 15: Loss curve of Convolutional Neural Network for Multi Class Classification

For CNN the training and validation loss starts from around 14 and gradually decreases to global minima of around 1. We are also not seeing any spikes or fluctuation in the curve. Moreover, there is also no sign of overfitting.

We also have compared our results with one of the research works [15] which is discussed in related works. Below is the comparison:

| Model Name | Papers | Testing Accuracy | Precision | Recall | F1 Score |
|---|---|---|---|---|---|
| XG Boost | Paper [15] | 98.89 | 98.93 | 98.89 | 98.89 |
| | Ours | 99.42 | 99 | 99 | 99 |
| Random Forest | Paper [15] | 98.54 | 98.54 | 98.5 | 98.5 |
| | Ours | 92.51 | 94 | 93 | 93 |
| CNN | Paper [15] | N/A | N/A | N/A | N/A |
| | Ours | 76.29 | 93.50 | 53.33 | 67.90 |
| ANN | Paper [15] | N/A | N/A | N/A | N/A |
| | Ours | 77.50 | 95.20 | 59.20 | 73 |

Table 2: Model Comparison Table for Multi Class Classification

From the above table we are seeing that for XGBoost model our model got 99.42%, 99%, 99% and 99% Accuracy, Precision, Recall and F1 Score respectively whereas the same model from another paper got 98.89%, 98.93%, 98.89% and 98.89% Accuracy, Precision, Recall and F1 Score respectively.

On the other hand, for Random Forest our model got 92.51%, 94%, 93% and 93% Accuracy, Precision, Recall and F1 Score respectively whereas the same model from another paper got 98.54%, 98.54%, 98.5% and 98.5% Accuracy, Precision, Recall and F1 Score respectively.

Furthermore, we are seeing that in our work we had applied CNN and ANN models, but we didn't find any data regarding the ANN and CNN model in the other paper.

## 5.4 Chapter Conclusion

After all the analysis in conclusion we can say that in our work Extreme Gradient Boosting (XGBoost) is working better than other three models and also it is finishing the training process faster than other models because it is taking around 30 epochs to do it whereas ANN and CNN model took more epochs to finish the training even if we are using early stopping. On the other hand, from the comparison table we are seeing that our XGBoost model is working better than the model from the other paper because our Accuracy, Precision, Recall and F1 Score of this model are 0.53%, 0.07%, 0.11%, 0.11% higher than the model from the other paper which is a significant improvement.

# CHAPTER 6

# BINARY AND MULTI CLASS CLASSIFICATION USING HYBRID MODELS

## 6.1 Chapter Background

In this part we will discuss the tools applied for binary and multi class classification with hybrid models. We have integrated different machine learning models into two different hybrid models where one is for binary, and another one is for multi class classification. Python is the programming language which we have picked since the libraries are too abundant for this type of task and it is simple to operate. For the libraries we had consulted to create and train the models Scikit Learn.

In the next sections we will talk about these methods and the results we have got.

## 6.2 Methodologies

We had proposed 2 different hybrid models which use voting for classification. In these two models we had combined XGBoost, RF, SVM and KNN for binary classification and had combined RF, XGBoost and AdaBoost for multi class classification. In the next sections we will talk briefly about these methods.

### 6.2.1 Random Forest

Random Forest is one of several ensemble methods of machine learning that offer a versatile approach, suitable for both regression and classification. The mode of the classes used for classification tasks can be generated by training with a large number of decision trees. This method makes random forests quite successful for both classification and regression project and helps to lower overfitting [35].

**6.2.2 Extreme Gradient Boosting (XGBoost)**

Considered a subset of ensemble learning and a machine learning method included into the gradient boosting framework is extreme gradient boost boosting (XGBoost). Beginning with simple learners, decision trees regularization methods help to increase model generalization. XGBoost is clearly computationally efficient; it offers good processing, perceptive feature significance analysis, and smooth management of missing data. In a study Chen et al. said that "Especially in problems with complicated data linkages, XGBoost can reach state-of-the-art performance by means of this sequential boosting with regularization approach." [36]

**6.2.3 K-Nearest Neighbor**

Using a distance metric, typically Euclidean distance, the KNN algorithm classifies data points according to the majority label of their nearest neighbors; it finds extensive use in non-parametric machine learning tasks like regression and classification. Although KNN is computationally heavy for big datasets since it must compute distances for every query point, it is popular for its ease of use and effectiveness in uses including pattern detection and recommendation systems [37].

**6.2.4 Support Vector Machine**

SVM is a supervised ML method for regression and classification tasks, typically with a focus on classification. SVM improves the generalizing capacity of a model by finding an ideal hyperplane that maximizes data points from many classes. This is achieved by maximizing the margin, that is, the distance separating the hyperplane from the closest support vector from every class [38].

**6.2.5 AdaBoost**

Combining several weak classifiers, the AdaBoost (adaptive boost) algorithm is an ensemble learning method producing a strong classifier with better accuracy. AdaBoost strengthens the general prediction performance of the model by iteratively changing the weights of training data, hence focusing on the samples that past classifier misclassified. This method especially helps to raise classification accuracy in many different fields [39].

**6.2.6 Hybrid Voting Classifier**

In machine learning, a hybrid model is combining several methods or models to improve predicted accuracy and robustness which is based on voting. In these hybrid systems, voting functions as a consensus process whereby every model vote on the forecast result, and the final choice is determined depending on a weighted or majority voting system. Particularly useful in classification problems, voting-based hybrid algorithms use the different capabilities of several models to manage heterogeneous and complex data, hence improving prediction accuracy [40].

Now we will talk about the evaluation metrics and results.

**6.3 Evaluation and Discussion**

We have evaluated my models based on 4 different scales such as F1-Score, Accuracy, Precision and Recall. Below are some brief descriptions of those metrics:

**6.3.1 Accuracy**

In ML or DL, accuracy is a performance indicator that illustrates how many correct predictions it generates.

### 6.3.2 Precision

Precision is defined as the ratio of true positive predictions to the total number of positive cases projected.

### 6.3.3 Recall

Recall is another machine learning or deep learning statistic gauging a model's accuracy in recognizing positive occurrences; sensitivity or true positive rate is another name for recall.

### 6.3.4 F1-Score

Combining accuracy and recall, the F1-score presents a single assessment of the performance of a model. Presenting a decent balance between the two metrics, memory's harmonic mean is accuracy's counterpart.

Now we will discuss the bar graphs of these metrics for my binary and multi class classification hybrid models.

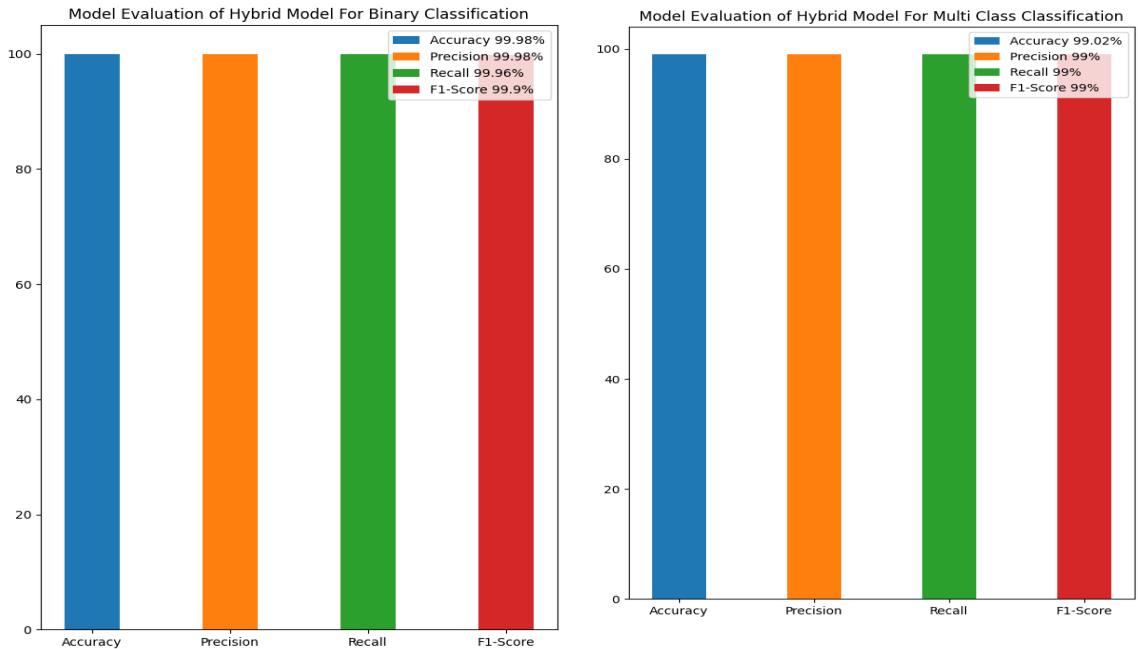

Fig 16: Evaluation of Hybrid Models for Binary and Multi Class Classification

From the bar graph of model evaluation for binary classification showing that the Accuracy, Precision, Recall and F1-Score are 99.98%, 99.98%, 99.96% and 99.9% respectively which is very high. On the other hand, the bar graph of model evaluation for multi class classification showing that the average Accuracy, Precision, Recall and F1-Score are 99.02%, 99%, 99% and 99% respectively which is also high.

Now we will talk about the accuracy curve of these two hybrid models.

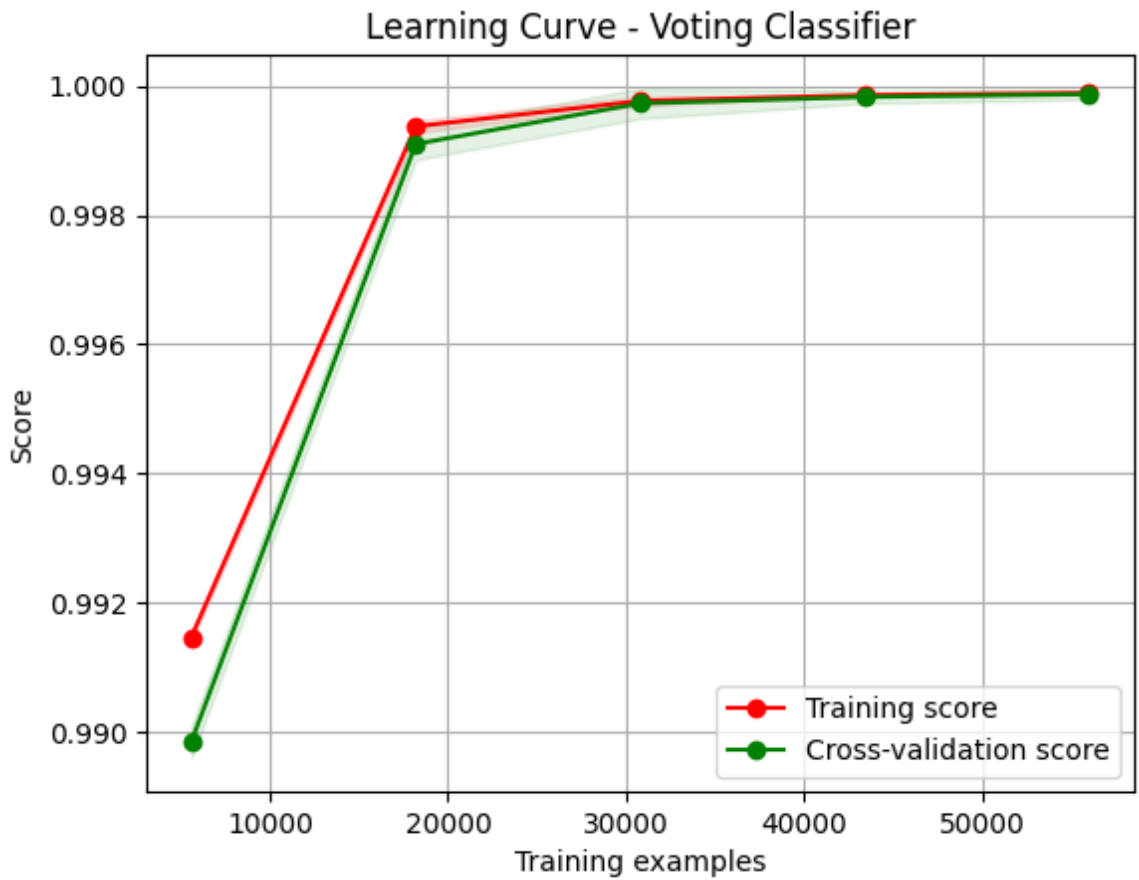

Fig 17: Hybrid Voting Classifier Accuracy Curve for Binary Classification

From the above accuracy curve of hybrid model for binary classification showing that, the training and validation accuracy started from around 99% to 99.2% and it gradually increased to around 99.9%. It also shows that there is no spike in the graph so both the validation and training curve are converging smoothly. Moreover, the validation accuracy curve is not going downwards at any point so there is no sign of model overfitting.

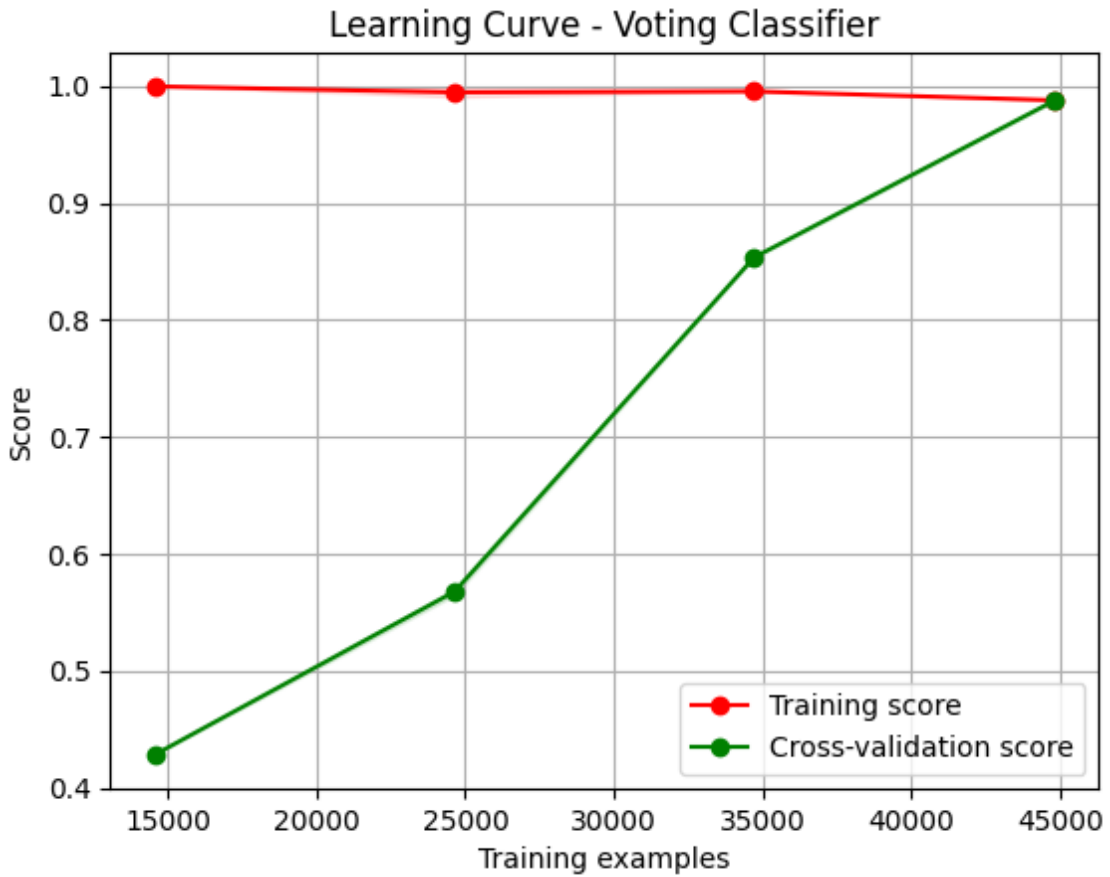

Fig 18: Hybrid Voting Classifier Accuracy Curve for Multi Class Classification

The above accuracy curve for the multi class classification shows that the validation accuracy started from around 40% but the training accuracy started from 99%. Gradually after so many training iterations the validation slowly climbs to the accuracy of around 99%. The graph also shows that there is no spike in the validation or training curve, so it is converging smoothly, also, there is no point when the validation accuracy started to decrease comparing to the training accuracy so there is no sign of model overfitting. Now we will compare these models with two of the research work [5] [15] which was described briefly in the literature review.

| Model Name | Papers | Testing Accuracy | Precision | Recall | F1 Score |
|---|---|---|---|---|---|
| Random Forest | Paper [5] | 95 | 59 | 44 | 50 |
| Hybrid | Ours | 99.98 | 99.98 | 99.96 | 99.9 |

Table 3: Comparison Table of the Hybrid Model for Binary Classification

| Model Name | Papers | Testing Accuracy | Precision | Recall | F1 Score |
|---|---|---|---|---|---|
| XG Boost | Paper [15] | 98.89 | 98.93 | 98.89 | 98.89 |
| Hybrid | Ours | 99.02 | 99 | 99 | 99 |

Table 4: Comparison Table of the Hybrid Model for Multi Class Classification

The first comparison table of binary classification showing that in the paper [5] that we choose to compare my hybrid model worked with standalone random forest model and got Accuracy, Precision, Recall and F1-Score of 95%, 59%, 44% and 50% respectively and this was the best model of that work but from my hybrid model we got Accuracy, Precision, Recall and F1-Score of 99.98%, 99.98%, 99.96% and 99.9% respectively which is 4.98%, 40.98%, 55.96% and 49.9% higher respectively on those evaluation metrics.

On the other hand, the second comparison table of multi class classification showing that, in paper [15] they had chosen XGBoost as their best model and got Accuracy, Precision, Recall and F1-Score of 98.89%, 98.93%, 98.89% and 98,89% respectively but from my hybrid model we got Accuracy, Precision, Recall and F1-Score of 99.02%, 99%, 99% and 99% respectively which is 0.13%, 0.07%, 0.07% and 0.07% higher respectively on those evaluation metrics.

**6.4 Chapter Conclusion**

After analyzing the bar graphs of the evaluation metrics and the comparison table in conclusion we can say that my hybrid models for multi class and binary classification are performing better than the standalone models which is used for comparison. Moreover, standalone models have some limitations for which sometimes they don't work better for complex dataset like IoT23 and as hybrid models combine the power of multiple models which is crucial for complex dataset so from this perspective, we can say that using hybrid models is a better approach for tackling classification tasks using complex datasets.

# CHAPTER 7

# CONCLUSION AND FUTURE WORKS

**7.1 Conclusion**

This article examined the application of deep learning and machine learning models with multi-class and binary classifiers for intrusion detection in IoT networks using the IoT23 dataset. Standalone ML models such as Random Forest, XGBoost, K-Nearest Neighbors, Support Vector Machine, and standalone deep learning models such as Convolutional Neural Network and Artificial Neural Network were evaluated for the performance. Furthermore, two voting based hybrid models have also been created. For binary classification K-Nearest Neighbor, XGBoost, Random Forest, and Support Vector Machine were combined; and for multi class classification, XGBoost, RF, and AdaBoost were combined.

Lastly, from the observed results it is deduced that for binary classification XGBoost is giving 98.9%, 98.5%, 98.7% and 98.9% Accuracy, Precision, Recall and F1-Score, respectively and for multi class classification it is giving 99.42%, 99%, 99%, 99% of Accuracy, Precision, Recall and F1-Score, respectively which is better than the models used for comparison from paper [5] and [15]. On the other hand, for hybrid models, they are giving 99.98%, 99.98%, 99.96% and 99.9% Accuracy, Precision, Recall and F1-Score, respectively for binary classification and for multi-class classification it is giving 99.02%, 99%, 99%, 99% Accuracy, Precision, Recall and F1-Score, respectively, which are also better than the models from those reference papers which are mentioned before. So, as a summary Extreme Gradient Boosting (XGBoost) and hybrid models for binary and multi-class classification are the best models for this classification task.

**7.2 Future Work**

Several paths of future research are suggested to progress the capabilities of intrusion detection systems (IDS) in Internet of Things networks. First of all, hyperparameter tuning and feature engineering show great potential to maximize model performance. Deep learning (DL) and machine learning (ML) models could benefit from advanced feature selection approaches like particle swarm optimization or evolutionary algorithms by enhancing signal clarity and decreasing noise. For resource-constrained Internet of Things (IoT) systems to provide real-time intrusion detection, these enhancements may be required to generate models with higher computational efficiency.

Secondly, it might be investigated to improve the performance of the hybrid model by means of advanced ensemble methods such weighted voting and stacking. In multi-class intrusion detection activities, ensembles can perhaps attain better accuracy and reduced false positive rates by giving model-specific weights depending on performance criteria. Adaptive learning models and semi-supervised learning techniques are still under investigation in another sense. These methods let models constantly change to fit new kinds of attacks, hence strengthening resilience against changing cyberthreats. Without requiring large-scale labeled data, semi-supervised learning and reinforcement learning approaches could help IDS to better detect abnormalities or unknown attack paths. Furthermore, as the field of IoT is quite broad and different devices have varied behaviors thus, the implementation of transfer learning will enable a model which can be trained in one dataset and can be adopted in different datasets or situations.

Finally, implementing these models in actual Internet of Things systems would offer insightful analysis of their scalability and useful efficacy. Testing in several IoT configurations, including smart cities, industrial IoT, and home automation, could draw attention to model resilience and flexibility in many contexts. Real-world implementation would also help research of computational trade-offs and possible model improvements required for high performance in Internet of Things systems.

# BIBLIOGRAPHY

[1] D. Upadhyay, J. Manero, M. Zaman and S. Sampalli, "Intrusion Detection in SCADA Based Power Grids: Recursive Feature Elimination Model with Majority Vote Ensemble Algorithm," in IEEE Transactions on Network Science and Engineering, vol. 8, no. 3, pp. 2559-2574, 1 July-Sept. 2021, doi: 10.1109/TNSE.2021.3099371.

[2] Cao B, Li C, Song Y, Qin Y, Chen C. Network Intrusion Detection Model Based on CNN and GRU. *Applied Sciences*. 2022; 12(9):4184. https://doi.org/10.3390/app12094184

[3] G. Guo, "An Intrusion Detection System for the Internet of Things Using Machine Learning Models," *2022 3rd International Conference on Big Data, Artificial Intelligence and Internet of Things Engineering (ICBAIE)*, Xi'an, China, 2022, pp. 332-335, doi: 10.1109/ICBAIE56435.2022.9985800.

[4] Churcher A, Ullah R, Ahmad J, Ur Rehman S, Masood F, Gogate M, Alqahtani F, Nour B, Buchanan WJ. An Experimental Analysis of Attack Classification Using Machine Learning in IoT Networks. *Sensors*. 2021; 21(2):446. https://doi.org/10.3390/s21020446

[5] G. Bhandari, A. Lyth, A. Shalaginov, and T.-M. Grønli, "Distributed deep neural-network-based middleware for cyber-attacks detection in smart IoT ecosystem: A novel framework and performance evaluation approach," Electronics, vol. 12, no. 2, p. 298, 2023.

[6] M. V. R. Sarobin, J. Ranjith, D. Ashwath, K. Vinithi, V. Khushi et al., "Comparative analysis of various feature extraction methods on IoT 2023," Procedia Computer Science, vol. 233, pp. 670–681, 2024.

[7] Ullah and Q. H. Mahmoud, "Design and development of a deep learning-based model for anomaly detection in IoT networks," IEEE Access, vol. 9, pp. 103906–103926, 2021.

[8] Malele, V., & Mathonsi, T. (2023). Testing the performance of Multi-class IDS public dataset using Supervised Machine Learning Algorithms. *ArXiv*, abs/2302.14374. https://doi.org/10.48550/arXiv.2302.14374.

[9] Banadaki, Y. (2020). Evaluating the performance of machine learning algorithms for network intrusion detection systems in the internet of things infrastructure. *Journal of Advanced Computer Science & Technology*. https://doi.org/10.14419/JACST.V9I1.30992.

[10] Jiang, H., He, Z., Ye, G., & Zhang, H. (2020). Network Intrusion Detection Based on PSO-Xgboost Model. *IEEE Access*, 8, 58392-58401. https://doi.org/10.1109/ACCESS.2020.2982418.

[11]     Churcher, A., Ullah, R., Ahmad, J., Rehman, S., Masood, F., Gogate, M., Alqahtani, F., Nour, B., & Buchanan, W. (2021). An Experimental Analysis of Attack Classification Using Machine Learning in IoT Networks. *Sensors (Basel, Switzerland)*, 21. https://doi.org/10.3390/s21020446.

[12]     Aljumah, A. (2021). IoT-based intrusion detection system using convolution neural networks. *PeerJ Computer Science*, 7. https://doi.org/10.7717/peerj-cs.721.

[13]     Cao, B., Li, C., Song, Y., Qin, Y., & Chen, C. (2022). Network Intrusion Detection Model Based on CNN and GRU. *Applied Sciences*. https://doi.org/10.3390/app12094184.

[14]     Hussein, S., Mahmood, A., & Oraby, E. (2021). Network Intrusion Detection System Using Ensemble Learning Approaches. *Webology*. https://doi.org/10.14704/web/v18si05/web18274.

[15]     Alrefaei A, Ilyas M. Using Machine Learning Multiclass Classification Technique to Detect IoT Attacks in Real Time. *Sensors*. 2024; 24(14):4516. https://doi.org/10.3390/s24144516

[16]     R. Liu, Z. Chen and J. Liu, "A Hybrid Intrusion Detection System Based on Feature Selection and Voting Classifier," *2023 IEEE 47th Annual Computers, Software, and Applications Conference (COMPSAC)*, Torino, Italy, 2023, pp. 203-212, DOI: 10.1109/COMPSAC57700.2023.00034.

[17]     Mhawi, D., Aldallal, A., & Hassan, S. (2022). Advanced Feature-Selection-Based Hybrid Ensemble Learning Algorithms for Network Intrusion Detection Systems. *Symmetry*, 14, 1461. https://doi.org/10.3390/sym14071461.

[18] Leevy, J., Hancock, J., Khoshgoftaar, T., & Peterson, J. (2021). Detecting Information Theft Attacks in the Bot-IoT Dataset. *2021 20th IEEE International Conference on Machine Learning and Applications (ICMLA)*, 807-812. https://doi.org/10.1109/ICMLA52953.2021.00133.

[19] Derhab, A., Aldweesh, A., Emam, A., & Khan, F. (2020). Intrusion Detection System for Internet of Things Based on Temporal Convolution Neural Network and Efficient Feature Engineering. *Wirel. Commun. Mob. Comput.*, 2020, 6689134:1-6689134:16. https://doi.org/10.1155/2020/6689134.

[20] N.-A. Stoian, "Machine learning for anomaly detection in IoT networks: Malware analysis on the IoT-23 dataset," B.S. thesis, University of Twente, 2020.

[21] Mazumder, A., Kamruzzaman, N., Akter, N., Arbe, N., & Rahman, M. (2021). Network Intrusion Detection Using Hybrid Machine Learning Model. *2021 International Conference on Advances in Electrical, Computing, Communication and Sustainable Technologies (ICAECT)*, 1-8. https://doi.org/10.1109/ICAECT49130.2021.9392483.

[22] Ozsahin, D., Mustapha, M., Mubarak, A., Ameen, Z., & Uzun, B. (2022). Impact of feature scaling on machine learning models for the diagnosis of diabetes. *2022 International Conference on Artificial Intelligence in Everything (AIE)*, 87-94. https://doi.org/10.1109/aie57029.2022.00024.

[23]     Butun, S. D. Morgera, and R. Sankar, "A survey of intrusion detection systems in wireless sensor networks," IEEE communications surveys & tutorials, vol. 16, no. 1, pp. 266–282, 2013.

[24]     Bai, Y. (2022). RELU-Function and Derived Function Review. *SHS Web of Conferences*. https://doi.org/10.1051/shsconf/202214402006.

[25]     Goel, S., Sharma, S., & Tripathi, R. (2021). Predicting Diabetes using CNN for Various Activation Functions: A Comparative Study. *2021 10th International Conference on System Modeling & Advancement in Research Trends (SMART)*, 665-669. https://doi.org/10.1109/SMART52563.2021.9676280.

[26]     Jap, D., Won, Y., & Bhasin, S. (2021). Fault injection attacks on SoftMax function in deep neural networks. *Proceedings of the 18th ACM International Conference on Computing Frontiers*. https://doi.org/10.1145/3457388.3458870.

[27]     Jiang, H., Tao, C., Dong, Y., & Xiong, R. (2020). Robust low-rank multiple kernel learning with compound regularization. *Eur. J. Oper. Res.*, 295, 634-647. https://doi.org/10.1016/j.ejor.2020.12.024.

[28]     Boschi, T., Reimherr, M.L., & Chiaromonte, F. (2020). An Efficient Semi-smooth Newton Augmented Lagrangian Method for Elastic Net. *ArXiv, abs/2006.03970*.

[29]     Wang, Q., Ma, Y., Zhao, K., & Tian, Y. (2020). A Comprehensive Survey of Loss Functions in Machine Learning. *Annals of Data Science*, 1-26. https://doi.org/10.1007/s40745-020-00253-5.

[30]     Chen, C., Lin, P., Hsieh, J., Cheng, S., & Jeng, J. (2020). Robust Multi-Class Classification Using Linearly Scored Categorical Cross-Entropy. *202020 3rd IEEE International Conference on Knowledge Innovation and Invention (ICKII)*, 200-203. https://doi.org/10.1109/ICKII50300.2020.9318835.

[31]     Ebrahim, M., Sedky, A., & Mesbah, S. (2023). Accuracy Assessment of Machine Learning Algorithms Used to Predict Breast Cancer. *Data*, 8, 35. https://doi.org/10.3390/data8020035.

[32]     Erkan, U. (2020). A precise and stable machine learning algorithm: eigenvalue classification (EigenClass). *Neural Computing and Applications*, 33, 5381-5392. https://doi.org/10.1007/S00521-020-05343-2.

[33]     Chung, J., & Lee, K. (2023). Credit Card Fraud Detection: An Improved Strategy for High Recall Using KNN, LDA, and Linear Regression. *Sensors (Basel, Switzerland)*, 23. https://doi.org/10.3390/s23187788.

[34]     Takahashi, K., Yamamoto, K., Kuchiba, A. *et al.* Confidence interval for micro-averaged $F_1$ and macro-averaged $F_1$ scores. *Appl Intell* **52**, 4961–4972 (2022). https://doi.org/10.1007/s10489-021-02635-5

[35]     Schonlau, M., & Zou, R. (2020). The random forest algorithm for statistical learning. *The Stata Journal*, 20, 29 - 3.

[36]     Chen, J., Zhao, F., Sun, Y., & Yin, Y. (2020). Improved XGBoost model based on genetic algorithm. *Int. J. Comput. Appl. Technol.*, 62, 240-245. https://doi.org/10.1504/ijcat.2020.10028423.

[37]     Wang, H., Xu, P., & Zhao, J. (2021). Improved KNN Algorithm Based on Preprocessing of Center in Smart Cities. *Complex.*, 2021, 5524388:1-5524388:10. https://doi.org/10.1155/2021/5524388.

[38]     Abdullah, D., & Abdulazeez, A. (2021). Machine Learning Applications based on SVM Classification A Review. *Qubahan Academic Journal*. https://doi.org/10.48161/QAJ.V1N2A50.

[39]     Li, T., Chen, X., & Li, W. (2023). An Improved AdaBoost Method in Imbalanced Data Learning. *2023 International Conference on Cyber-Physical Social Intelligence (ICCSI)*, 405-410.

[40]     Du, S., Ding, W., Yang, D., & Yang, L. (2022). Research on machine learning strategy based on voting model. *Conference on Computer Science and Communication Technology*.

[41]     Prazeres, N., Costa, R. L. C., Santos, L., & Rabadão, C. (2022). Evaluation of AI-based Malware Detection in IoT Network Traffic. In *SECRYPT* (pp. 580-585).